\documentclass{article}
\usepackage[utf8]{inputenc}
\usepackage{authblk}
\usepackage[margin=1.25in]{geometry}
\usepackage{graphicx}
\graphicspath{ {./figures/} }
\usepackage{subcaption}
\usepackage{amsmath}
\usepackage{amssymb} 
\usepackage{booktabs}
\usepackage{doi}
\usepackage{setspace}
\usepackage[T1]{fontenc}

\usepackage[round,authoryear]{natbib}
\usepackage{aas_macros}
\title{Transients and Stellar Activity as Revealed by Ultraviolet Diagnostic Tracers}
\author[1,2,3,*]{Richard de Grijs}
\author[4]{Martin A. Barstow}
\author[1,2,5]{Deepak Chahal}
\author[6,7,3]{Fabio Favata}
\author[8]{Xiaoting Fu}
\author[9]{Ana I. G\'omez de Castro}
\author[10,11]{Sipra Hota}
\author[10]{Firoza Sutaria}

\affil[1]{School of Mathematical and Physical Sciences, Macquarie University, Balaclava Road, Sydney, NSW 2109, Australia.}
\affil[2]{Astrophysics and Space Technologies Research Centre, Macquarie University, Balaclava Road, Sydney, NSW 2109, Australia.}
\affil[3]{International Space Science Institute--Beijing, 1 Nanertiao, Zhongguancun, Hai Dian District, Beijing 100190, China.}
\affil[4]{School of Physics \& Astronomy, University of Leicester, Leicester LE1 7RH, UK.}
\affil[5]{Astronomy Unit, Department of Physics and Astronomy, Queen Mary University of London, Mile End Road, London E1 4NS, UK.}
\affil[6]{INAF--Osservatorio Astronomico di Palermo, Piazza del Parlamento, 1, 90134 Palermo, Italy.}
\affil[7]{Department of Physics, Imperial College London, Exhibition Road, London SW7 2AZ, UK.}
\affil[8]{Purple Mountain Observatory, Chinese Academy of Sciences, Nanjing, Jiangsu 210023, China.}
\affil[9]{Space Astronomy Research Group--AEGORA, Facultad de Ciencias Matem\'aticas, Universidad Complutense de Madrid, Plaza de Ciencias 3, 28040 Madrid, Spain.}
\affil[10]{Indian Institute of Astrophysics, 2nd Block, Koramangala, Bangalore-560034, India.}
\affil[11]{Pondicherry University, R.V. Nagar, Kalapet, Puducherry-605014, India.}
\affil[*]{Address correspondence to: richard.de-grijs@mq.edu.au}

\date{2026}

\begin{document}

\maketitle

%%%%%% Abstract %%%%%%
\begin{abstract}
Ultraviolet (UV) observations offer unique insights into transient and variable astrophysical phenomena, from the Sun and nearby stars to supernovae and star-forming galaxies. Here, we summarise and review the diagnostic power of the UV spectrum for probing the physical conditions, dynamics and energetics of these events. We trace the historical development of UV astronomy from early satellite missions to the current landscape shaped by the Hubble Space Telescope, the GALaxy Evolution eXplorer, the Neil Gehrels Swift Observatory and {\sl AstroSat}/UVIT. Particular emphasis is placed on resonance lines such as Mg\,{\sc ii} $h$ and $k$, C\,{\sc ii}, Si\,{\sc iv}, C\,{\sc iv} and N\,{\sc v}, which probe the (solar and) stellar chromosphere and transition region, revealing magnetic heating, shocks and mass loss across stellar types. We examine UV diagnostics of stellar flares, pre-main-sequence activity and rotation--activity correlations, linking them to exoplanetary irradiation and habitability. We then explore the UV view of supernova shock breakout and early-time emission, as well as the far-UV mapping of young stellar populations and hierarchical star formation in the Magellanic Clouds. Together, these aspects illustrate how the UV connects small-scale plasma processes to galactic-scale star formation. Despite remarkable progress, UV transient astronomy remains limited by instrument availability and cadence. Forthcoming missions promise to transform this field by capturing the rapid, luminous phenomena that define the dynamic Universe.
\end{abstract}

%%%%%% Main Text %%%%%%

\section{Setting the scene}

From its inception, the exploration of the Universe at ultraviolet (UV) wavelengths has been driven by a dual desire: (i) to access the rich forest of atomic and ionic transitions that fall in this spectral range and (ii) to exploit these diagnostics to reveal the dynamic processes shaping stars and the galaxies they form. Here, we will adopt the conventional astronomical definitions of the UV regime. The UV broadly encompasses wavelengths from $\sim$100 to 4000 {\AA}, bridging the optical domain and the soft X-ray regime. For clarity, we refer to the near-UV as the interval from approximately 2000 to 4000 {\AA}, where chromospheric and upper-photospheric diagnostics dominate. The far-UV, spanning $\sim$912--2000 {\AA}, contains key resonance lines (e.g., C\,{\sc iv}, Si\,{\sc iv}, N\,{\sc v}) that trace transition-region plasma and radiatively driven winds. At still shorter wavelengths, the extreme-UV ($\sim$100--912 {\AA}) probes highly ionised species in hot ($10^{5-6}$ K) stellar atmospheres, although interstellar H\,{\sc i} absorption severely limits extragalactic access to this regime. 

These spectral subdivisions reflect both physical transitions in stellar atmospheres (e.g., 912 {\AA} corresponds to the hydrogen Lyman limit) and the instrumental capabilities of space-borne UV observatories. In the context of these definitions, for the Sun the upper photosphere corresponds to our star's visible surface, where temperatures are slightly cooler than in the deeper photosphere, roughly 4500--6000 K. The gas density is lower than in deeper photospheric layers, so the plasma is more transparent. The upper photosphere encompasses the first $\sim$500 km above the solar surface. The chromosphere is located just above the photosphere. It is less dense than the photosphere, with temperatures rising from $\sim$6000 K at the bottom (some 500 km above the solar surface) to $\sim$20,000~K at the top, at an altitude of around 2000 km. On top of the chromosphere, the `transition region' to the tenuous but extremely hot (1--5 MK) corona is very thin, just a few hundred km, but the temperature jumps steeply from $\sim$20,000 K to $\sim$500,000 K (see also Section \ref{sec2}). 

Because the Earth's atmosphere blocks nearly all UV photons, the field's origin coincided with the early Space Age. It has since developed into a cornerstone of stellar astrophysics research, with UV spectroscopy now providing some of the best-constrained tracers of stellar atmospheres, activity and evolution. Here we explore how the UV regime plays a pivotal role in connecting the microphysics of stellar surfaces and coronae to the macroscopic phenomena of stellar variability, explosive transients and even the star-formation histories of galaxies.

The earliest inroads into harnessing UV astronomy were necessarily modest. Suborbital rocket experiments in the 1950s and 1960s, led by pioneering groups in the USA and Europe, delivered the first glimpses of hot-star continua and interstellar absorption features shortwards of 300 nm \citep{1966ApJ...144....1M}. These early spectra demonstrated both the exciting opportunities and the serious challenges associated with UV observations, including, on the one hand, the availability of a wealth of resonance transitions from abundant ions such as C\,{\sc ii}, Si\,{\sc iv} and Mg\,{\sc ii}; on the other, the technical difficulty of developing stable detectors and reliable calibration above the Earth's atmosphere. Still, these pathfinders laid the foundation for the more ambitious satellite missions that followed, such as the Orbiting Astronomical Observatories ({\sl OAO}) in the late 1960s and early 1970s \citep{1970ApJ...161..377C}. {\sl OAO}-2, in particular, showed how UV photometry and low-resolution spectroscopy could be deployed systematically, revealing variations in stellar fluxes and the importance of hot stars in shaping galactic radiation fields \citep[e.g.,][]{1970ApJ...161..377C,1979ApJS...39..195C}.

A decisive advance came with {\sl Copernicus} ({\sl OAO}-3), launched in August 1972, which delivered high-resolution stellar UV spectroscopy of unprecedented quality \citep{1973ApJ...181L..97R}. It opened the door to quantitative studies of interstellar absorption, stellar winds and stellar chromospheric activity. Next, launched in January 1978, the International Ultraviolet Explorer ({\sl IUE}) revolutionised the field by offering a dedicated, long-lived platform for UV spectroscopy. During its operational lifetime (until September 1996), {\sl IUE} collected an enormous archive of UV spectra of both stars and galaxies, its modest aperture compensated by continuous accessibility. For the study of stellar activity (see Section \ref{sec2}), {\sl IUE} became a workhorse: the satellite traced variability in chromospheric emission lines, captured flare events and extended solar-inspired diagnostics to large stellar samples \citep{1978Natur.275..372B,1985ApJ...298..761B}.

\subsection{The solar connection}

Although much of the initial UV exploration was motivated by the astrophysics of hot stars and distant galaxies, the Sun quickly emerged as a natural proving ground. The Sun's proximity made it the only object whose UV radiation could be spatially resolved, at least initially, thus allowing direct tests of theoretical ideas about chromospheric and coronal heating. In the 1960s, early rocket-borne measurements revealed the richness of the solar UV spectrum, with resonance lines such as C\,{\sc ii}, Si\,{\sc iv}, and Mg\,{\sc ii} $h$ and $k$ standing out as key diagnostics of temperature, density and velocity fields in the chromosphere and the transition region \citep{1981ApJS...45..635V}. These features soon became key diagnostics for interpreting stellar atmospheres more generally, thus ensuring that solar physics and stellar astrophysics research evolved in tandem.

Subsequent satellite observatories deepened this connection. The Orbiting Solar Observatories ({\sl OSO} series) and {\sl Skylab}'s Apollo Telescope Mount provided the first continuous coverage of solar UV emission, documenting its variability across active regions and during flares \citep{1965AnAp...28..446N,1966AJ.....71..173N,1977ApOpt..16..870T}. Later missions such as the Solar and Heliospheric Observatory ({\sl SOHO}) and the Transition Region and Coronal Explorer ({\sl TRACE}) in the 1990s, and the Solar Dynamics Observatory ({\sl SDO}) in the second decade of the current century, refined the picture still further. Their observations demonstrated that solar UV variability is not only a tracer of local magnetic activity but also an important driver of space weather (the influence of solar activity on the near-Earth environment and technological systems), with implications for Earth's ionosphere (an upper atmospheric layer that is ionised by solar and cosmic radiation) and planetary atmospheres more generally \citep{2000JGR...10527195W, 2012SoPh..275....3P}. These insights underlined the dual role of solar UV studies: (i) as a key branch of heliophysics in their own right and (ii) as a foundation for interpreting the spectral fingerprints of other stars.

As such, the historical arc of UV astronomy is inseparable from the development of our understanding of stellar variability and transients. As instruments matured, so too did the ambition of the questions posed. By the 1980s and 1990s, attention had shifted from cataloguing spectral features to tracking how they varied in time: stellar flares on active dwarfs, mass-loss episodes from evolved giants, pulsation cycles in Cepheids and the early-time signatures of supernovae (SNe). The time-domain dimension of UV astronomy has always been technically challenging---short exposures, limited sensitivity and the need for rapid response---but it became increasingly central to the field's predominant research direction. Indeed, the same chromospheric (magnetic) resonance lines---Mg\,{\sc ii}, C\,{\sc ii}, Si\,{\sc iv} and C\,{\sc iv}---that trace plages (bright patches in the chromosphere) and prominences (magnetic filaments) on the Sun also flare dramatically in cool stars, thus forging a powerful link between local solar physics and the broader behaviour of stellar populations \citep[][see Section \ref{sec5} for a discussion of UV solar and stellar flares]{1988A&A...206L...1B, 2014ApJS..211....9L}.

The launch of the Hubble Space Telescope ({\sl HST}) in April 1990---equipped with instruments such as the Goddard High Resolution Spectrograph (GHRS) and, later, the Space Telescope Imaging Spectrograph (STIS) and the Cosmic Origins Spectrograph (COS)---brought a new level of spectral precision to UV diagnostics. With the {\sl HST}, it became possible to measure line profiles in exquisite detail and probe velocity fields, atmospheric turbulence and absorption structures in both stellar and interstellar contexts \citep[e.g.,][]{1995ApJ...444..424C}. Complementing {\sl HST}'s near- and far-UV coverage, the Far Ultraviolet Spectroscopic Explorer ({\sl FUSE}; 1999--2007) provided high-resolution ($R \equiv \frac{\lambda}{\Delta \lambda} \sim 20,000$) spectroscopy at wavelengths inaccessible to {\sl HST}, particularly in the 905--1187 {\AA} range. It enabled transformative studies of O\,{\sc vi}, the hydrogen Lyman series and molecular hydrogen in hot stellar atmospheres and the interstellar medium \citep[ISM; e.g.,][]{2000ApJ...538L...1M,2000ApJ...538L...7S}. Together, these facilities redefined UV diagnostics across stellar and interstellar environments. {\sl HST}’s UV instruments, in particular, were critical in capturing the brief, luminous UV flashes that mark the shock breakout (SBO) of core-collapse SNe (see Section \ref{sec6}), as well as for mapping accretion phenomena in cataclysmic variables and young stellar objects. The temporal resolution and sensitivity now allowed transients to be caught in the act rather than inferred retrospectively.

At the same time, dedicated missions such as the GAlaxy Evolution eXplorer \citep[{\sl GALEX}, launched in April 2003;][]{2014..Simons,2017ApJS..230...24B} extended UV studies into the wide-field extragalactic domain, mapping star formation across nearby galaxies (although at low spatial resolution) and providing context for the stellar activity seen in the Milky Way \citep{2005ApJ...619L...1M}. {\sl GALEX} demonstrated the power of UV tracers on galactic scales, whereas smaller missions such as the Neil Gehrels Swift Observatory \citep[{\sl Swift};][]{2017MNRAS.466.4540H,2025..Ludwig}, with its UV/Optical Telescope (UVOT), kept time-domain monitoring of SNe and gamma-ray bursts firmly in view \citep{2005SSRv..120...95R}. Together, these observatories established a continuum of sorts, from the physics of stellar atmospheres to the demographics of galaxies.

From a theoretical perspective, UV diagnostics have always occupied a privileged place because they access physical regimes that are otherwise inaccessible. Chromospheric resonance lines form at temperatures where non-local thermodynamic equilibrium (non-LTE) conditions dominate, making them sensitive to departures from radiative equilibrium and to the influence of magnetic heating \citep{1978ApJ...220..619L, 1991A&A...252..203R}. Transition-region lines trace plasma at temperatures of approximately $10^{4-6}$ K, thus providing unique diagnostics of flare energetics and coronal heating processes \citep{2015A&A...583A..54S, 2018LRSP...15....5D}. At higher energies, UV continua and the effects of non-LTE line blanketing (the cumulative effect of many absorption lines in a star’s spectrum on its emergent radiation) in hot stars shape both photospheric opacities and serve as key signatures of strong, radiatively driven winds \citep{1975ApJ...195..157C, 1998ApJ...496..407H, 2023IAUS..370..130S}. In all these cases, the UV regime probes the boundaries---between photosphere and chromosphere, between steady outflow and eruptive mass loss, between quiescence and transient flares.

As the next sections explore in detail, the applications of UV tracers span an extraordinary range. For low-mass stars, they reveal the architecture of the stellar atmosphere from photosphere to corona, and they provide the best diagnostics of flare energetics and long-term activity cycles. For high-mass stars, they expose the unstable winds and their eventual demise as SNe, where the first photons to escape the envelope often appear in the UV. And beyond the Milky Way, UV spectroscopy and photometry offer the most direct probes of massive-star populations and star-formation histories, particularly in metal-poor environments such as those characteristic of our nearest neighbours, the Magellanic Clouds (see Section \ref{sec7}).

It is worth stressing that the transient and variable sky, now a dominant theme across modern astrophysics, has always had a natural affinity with UV diagnostics. The sudden heating of plasma, the emergence of shocks and the dispersal of circumstellar material are all processes that are observed most brightly at UV wavelengths. As new facilities emerge---both space-borne and ground-based proxies, such as near-UV imagers---the importance of UV tracers will only increase. They remain indispensable, not only because they provide unique diagnostics but also because they remind us that stellar activity and transients are not exotic outliers but rather the normal modes of a dynamic Universe.

\subsection{The UV as a diagnostic regime}
\label{uvdiagn.sec}

The UV domain contains numerous diagnostics of chromospheric, transition-region and coronal activity, including Mg\,{\sc ii}, C\,{\sc ii}, Si\,{\sc iv}, C\,{\sc iv} and N\,{\sc v}. These lines probe plasma spanning temperatures from $\sim 10^4$ K to $10^5$ K and provide sensitive tracers of magnetic heating, mass motions and atmospheric structure. Their formation and diagnostic interpretation are discussed in detail in Section \ref{sec2}.

UV observations are particularly sensitive to variability and transient behaviour. Solar and stellar flares, for instance, release most of their radiative energy in the UV, including the extreme UV \citep{1991ApJ...378..725H, 2019ApJ...871..167K}. 
%On M dwarfs, flare rates are so high that {\sl GALEX} time-domain surveys have catalogued thousands of events, revealing correlations between flare frequency, stellar rotation and magnetic activity \citep{2006A&A...458..921W}. These insights extend well beyond stellar physics. The UV radiation environment of exoplanet host stars, especially in the habitable zones of active M dwarfs, directly shapes planetary atmospheric photochemistry and habitability \citep{2016ApJ...820...89F}. Thus, UV activity tracers play a dual role: they diagnose stellar magnetism and also frame the conditions for planetary life.
Stellar flares represent one of the most dramatic manifestations of magnetic activity and are among the most important targets for UV observations. Their observational characteristics and physical interpretation are discussed in Section \ref{sec4}.

Although flare stars had long been recognised from ground-based optical surveys, particularly among dMe stars (flare-active M dwarfs) and BY Draconis (BY Dra)-type variables---cool, magnetically active M dwarfs whose optical spectra exhibit strong emission lines \citep{1972Ap&SS..19...75G})---the UV opened an entirely new window on their physics. {\sl IUE} provided the first opportunity to monitor stellar flares spectroscopically at high cadence. Observations of active late-type dwarfs revealed abrupt and large-amplitude enhancements in chromospheric and transition-region lines (most notably Mg\,{\sc ii} $h$ and $k$, C\,{\sc iv} and Si\,{\sc iv}), accompanied by a strong rise in UV continuum emission \citep{1981MNRAS.197..815B, 1991ApJ...378..725H}. These signatures demonstrated that flare heating can penetrate to atmospheric layers far above those accessible to optical diagnostics and that UV line intensities may exceed their quiescent levels by orders of magnitude. The simultaneous growth of optical Balmer and Ca\,{\sc ii} H and K emission with the UV continuum confirmed that the impulsive flare phase coincides with intense chromospheric heating and particle acceleration.

Subsequent space missions broadened both the scope and the diagnostic power of UV flare studies. {\sl GALEX} extended the investigation to a wider range of spectral types, showing that impulsive UV brightenings are common across the main sequence rather than confined to classical flare stars \citep{2007ApJS..173..673W}. {\sl HST}'s STIS and COS enabled time-resolved spectroscopy of individual flare events, revealing that multiple plasma components contribute to the UV response and that post-flare cooling persists long after the impulsive phase \citep{2014ApJS..211....9L}. {\sl Swift}, with its combined UV and X-ray sensitivity, has since provided a powerful means of correlating radiative outputs across bands, establishing the central role of magnetic reconnection (the process where magnetic-field lines snap and reconnect, releasing stored magnetic energy as heat, motion and energetic particles) in heating stellar coronae \citep{2016ApJ...832..174O, 2018ApJ...867...71L}. Collectively, these developments have shown that flares dominate the short-term variability of late-type stars in the UV and contribute substantially to their cumulative magnetic energy budget. The resulting UV radiation environments have profound implications for planetary atmospheres and for interpreting variability in stellar populations observed in the integrated light of galaxies.

In high-mass stars, the UV regime probes different physics. The strong resonance lines of C\,{\sc iv}, N\,{\sc v} and Si\,{\sc iv} reveal powerful stellar winds driven by line scattering of UV photons \citep{1975ApJ...195..157C, 1999isw..book.....L}. These winds influence stellar evolution, stripping mass and angular momentum and enriching the ISM with chemically processed material. The UV domain also captures the brief but luminous emission from SN SBOs. Models predict that the first light from core-collapse events emerges predominantly in the UV, with durations of minutes to hours \citep{1999ApJ...510..379M, 2017hsn..book..967W}. Detecting these UV flashes provides unique constraints on progenitor radii and circumstellar environments. The limited number of such detections to date underscores both the promise and the challenges of UV transient astronomy.

The reach of UV diagnostics extends well beyond the stars of our own Galaxy. In the Magellanic Clouds, where metallicities are well below solar, UV spectroscopy has been essential in testing how stellar winds depend on composition. Observations with {\sl HST} and {\sl FUSE} demonstrated that O-star winds in the Small Magellanic Cloud are weaker than their Galactic counterparts, consistent with the metallicity-dependent scaling predicted by line-driven wind theory \citep{2007A&A...473..603M}. More broadly, the rest-frame UV has become the cornerstone of extragalactic star-formation studies, particularly at high redshifts where optical observations probe the far-UV of young galaxies. Templates derived from UV spectroscopy of stars and star clusters in the Magellanic Clouds remain critical for interpreting galaxy spectra in the early Universe \citep[][see also Subramaniam et al., this volume]{1999ApJS..123....3L, 2010ApJS..189..309L, 2019ApJ...882..182C,2022ApJ...930..105S}. Thus, UV diagnostics form a continuous bridge: from stellar activity in the Sun and nearby stars, through massive-star winds and transients, to galaxy-scale star formation across cosmic time.

\subsection{The modern landscape of UV astronomy}
\label{modernUV}

Nevertheless, UV astronomy remains resource-limited. The dependence on space-borne platforms has meant that UV facilities are relatively few compared with ground-based optical or radio telescopes. {\sl HST} continues to operate, but its eventual retirement is approaching rapidly. {\sl GALEX} was decommissioned in 2013, and {\sl FUSE} ceased operations in 2007. Yet the future seems promising. Dedicated missions such as {\sl ULTRASAT}, scheduled for launch in the mid-2020s, promise to open the time-domain UV frontier with wide-field monitoring capable of capturing hundreds of early SN light curves each year \citep{2024ApJ...964...74S}. Proposed missions (for an extensive review, see Barstow et al., this volume) such as {\sl Polstar}, designed to combine UV spectroscopy with spectropolarimetry, would bring magnetic-field diagnostics into the UV \citep{2022Ap&SS.367..126S}. Meanwhile, multi-purpose observatories like the James Webb Space Telescope ({\sl JWST}), although lacking UV coverage, have reinforced the complementary role of the UV: {\sl JWST}'s infrared view of galaxies and exoplanets makes UV stellar data even more valuable as a counterpart. The synergy between UV and other wavelengths will only deepen as astrophysics moves further into the era of multi-messenger and time-domain exploration.

Several cross-cutting themes emerge from this narrative. First, the UV range provides high contrast between quiescent and active states. Where photospheric radiation dominates the optical regime, magnetic heating and transient processes stand out in the UV. Second, the UV unifies diagnostics across mass and evolutionary stage: the same wavelength regime that traces chromospheric heating in M dwarfs also reveals line-driven winds in O stars and the SBO flashes of SNe. Third, UV studies are fundamentally time-domain in character. The most revealing signals---whether flares, shocks or transient luminosity increases---are inherently variable, demanding instruments with both sensitivity and cadence. Finally, the UV connects disciplines: stellar astrophysics, exoplanet science, galactic evolution and cosmology all depend, in different ways, on UV diagnostics.

This introductory review section sets the stage for the more detailed discussions to follow. In Section \ref{sec2}, we turn to the theoretical underpinnings of stellar chromospheric lines and explain why UV resonance transitions serve as privileged diagnostics of activity. Section \ref{sec3} explores the lessons we can learn from variability studies of pre-main-sequence (PMS) stars. Section \ref{sec4} surveys the observational landscape for low-mass FGKM-type stars, emphasising the chromospheric, transition-region and coronal layers. Section \ref{sec5} focuses on UV flare diagnostics, whereas Section \ref{sec6} explores the UV view of massive stars on the brink of explosion, focussing on progenitor models and the fleeting UV flashes of SBO. Section \ref{sec7} extends our scope to extragalactic environments, showing how UV diagnostics trace star formation in the Magellanic Clouds and---ultimately---provide templates for high-redshift galaxies. Together, these sections build a comprehensive picture of how UV observations reveal stellar activity and transients across the cosmos.

\section{Stellar activity in the UV: Theoretical basis}
\label{sec2}

% Xiaoting Fu; focus on chromospheric lines
In this section, we will address the following three key questions, with a focus on the chromosphere and the transition region:
\begin{enumerate}
\item What does `stellar activity' mean in practice?
\item Why do UV lines trace stellar activity? and
\item How should we interpret key diagnostic lines tracing stellar activity?
\end{enumerate}

\subsection{What is stellar activity and why does the UV trace it?}

Building on the overview presented in Section \ref{uvdiagn.sec}, we now examine the physical basis of UV activity diagnostics in greater detail. `Stellar activity' is a catch-all term that refers to departures from purely radiative equilibrium and steady hydrostatic flows in the outer layers of stars. It includes additional heating, momentum input and time‐dependent dynamics occurring above the photosphere. The region of stellar activity includes the chromosphere, the transition region, the corona and any connected in- or outflows. Phenomenologically, activity reveals itself through enhanced emission over and above the photospheric continua, variability from seconds to decades, broadened or asymmetric spectral features, quasi‐periodic cycles and mass loss---ranging from transient ejections to steady winds. Physically, it is the conversion of mechanical, magnetic or gravitational energy into thermal energy, wave energy and bulk flows. On spatial scales, bulk flows range from flux‐tube widths to stellar radii, and on temporal scales they cover the time span from Alfv{\'e}n crossing times to magnetic and rotational cycles. Whereas magnetic fields are the dominant driver in many cool stars, stellar activity in a broader sense also includes pulsation‐driven shocks, radiation‐driven winds and accretion‐powered phenomena.

A basal (quiescent) chromospheric flux persists even in very inactive stars and is commonly attributed to non-magnetic heating \citep{Schroder2012}. It is driven by convective acoustic waves that steepen into shocks in the quiet chromosphere, independent of rotation. The magnetic excess above this floor correlates with rotation through the Rossby number \citep{Schrijver1987}, $Ro = P_\mathrm{rot}/\tau_\mathrm{conv}$, where $\tau_\mathrm{conv}$ is the convective turnover time---the characteristic timescale for a convective cell to rise to the surface---following the unsaturated $Ro$ scaling and then saturating for $Ro \lesssim 0.1$ at a nearly constant activity level.

In stars with a convective envelope and a relatively low effective temperature, including FGKM-type stars, stellar activity is predominantly magnetic. The interplay of convection, rotation and magnetic fields generates large amounts of magnetic energy. This energy is transported to the stellar surface and into the atmosphere, where it dissipates through processes like magnetic reconnection and wave heating, the process by which mechanical and magnetic waves generated in the star’s lower layers propagate upwards and deposit energy in the upper atmosphere. This non-radiative energy deposition heats the outer atmosphere to extreme temperatures. Above the photosphere, the chromosphere is characterised by an increasing temperature gradient and partial ionisation. It is followed by a narrow transition region at higher altitudes where the temperature and ionisation potential fraction rise steeply, leading into an overlying corona exhibiting high thermal and magnetic pressures. 

In hotter stars with radiative envelopes, including early A-, B- and O-type stars, large‐scale convective dynamos are weak or absent, and the dominant stellar activity arises from radiation forces, pulsation and accretion. Momentum is transferred from the stellar radiation field to the gas via metal-line opacity, in turn producing line‐driven winds whose UV resonance profiles (e.g., N\,{\sc v}, C\,{\sc iv}, Si\,{\sc iv}) show P Cygni shapes (reflecting strong outflows or stellar winds), discrete absorption components and clumping signatures. In radially pulsating stars (e.g., Cepheids, RR Lyrae and Mira variables), periodic shocks steepen in the atmosphere and heat/accelerate material; in accreting systems (young stars and interacting binary systems), gravitational energy released in accretion funnels or at boundary layers powers hotspots and fast flows. However, across these cases the drivers differ: radiation pressure, global pulsation dynamics or gravitational potential energy all form part of what we call stellar activity.

Given this context, one might wonder why diagnostic lines at UV wavelengths are especially responsive to stellar activity. Spectroscopically speaking, a line is deemed `diagnostic' when certain specific, measurable features (flux, flux ratio, width, shift, shape) map in a controlled fashion to a range of underlying plasma properties. The UV regime hosts a dense set of such probes that are selectively sensitive to the atmospheric layers where stellar activity originates.

Specific UV diagnostic lines are tuned to the thermodynamic regime where stellar activity is generated. A star's stratified outer atmosphere runs from the upper photosphere ($\sim\!6\times10^3$\,K) through the chromosphere and into the transition region, where the plasma is heated rapidly to $\sim\!10^{5-5.5}$\,K before it continues to feed the higher-level coronal plasma at $\gtrsim\!10^{6}$\,K.\footnote{Canonical temperature structure from semi-empirical solar models \citep{1981ApJS...45..635V,2009ApJ...707..482F}; coronal temperatures summarised by \cite{Reale2010,2014LRSP...11....4R}.} 
Many strong UV resonance lines arise from ions whose ionisation fractions and contribution functions peak in this temperature range, e.g., Si\,{\sc iv} ($\log T \mbox{ [K]}\!\approx\!4.9$), C\,{\sc iv} ($\log T \mbox{ [K]}\!\approx\!5.0$) and N\,{\sc v} ($\log T \mbox{ [K]}\!\approx\!5.3$).\footnote{Peak formation conditions (optically thin, ionisation equilibrium) from the {\sc chianti} database \citep{2016JPhB...49g4009Y,2019ApJS..241...22D,2021ApJ...909...38D}.} 
As such, any process that perturbs the heating rate, the pressure or the flows in these layers (magnetic reconnection, wave dissipation, pulsation shocks, line-driven instabilities, accretion impacts) directly changes the ionisation-level populations and emissivities of the UV transitions. Because the optical/near-UV photospheric continuum of cool stars is relatively bright, whereas the far-UV continuum is generally weak, the UV line-to-continuum contrast is high, thus rendering modest physical changes spectroscopically conspicuous.

Radiatively speaking, UV line formation is highly non-local and therefore very sensitive to activity. Densities in the chromosphere and transition region are low enough that collisional de-excitation rates are often small compared to radiative decay, so the line source function $S_\nu$ is dominated by scattering, i.e. $S_\nu \simeq (1-\epsilon)\,J_\nu + \epsilon\,B_\nu$, with the photon destruction probability $\epsilon\!\ll\!1$ and $B_\nu$ representing the Planck blackbody function (thermal emission). In this regime, the angularly averaged radiation field $J_\nu$, which carries information from neighbouring altitudes and structures, controls the emergent intensity. Thus, modest perturbations in temperature, density or velocity along the line of sight can imprint outsized changes onto UV line cores and wings. Line cores form at higher altitudes within the stellar atmosphere, where the optical depth is large, and respond to overlying cool/hot layers and flows; line wings originate deeper into the stellar atmosphere and track gradients and micro-turbulence. For a few resonance lines (Ly$\alpha$, Mg\,{\sc ii} $h$ and $k$), line wings also track partial frequency redistribution (PRD) physics, which describes how the frequency of a photon changes when it is absorbed and re-emitted.

Kinematically speaking, UV lines offer clean, fast probes of dynamics occurring on activity timescales. Their large oscillator strengths and short radiative lifetimes produce sharp, high-contrast profiles where Doppler shifts of a few km\,s$^{-1}$ in the quiescent transition region \citep{Chen2022} to hundreds of km\,s$^{-1}$ in winds, accretion events and flares \citep{Prinja1990}. Bulk flows associated with chromospheric evaporation/condensation, pulsation-driven shocks, accretion funnels or wind acceleration appear as asymmetric wings, centroid shifts or P Cygni-type morphologies. Non-thermal broadening (from waves and turbulence) adds to thermal line widths and grows with formation temperature, thus enabling a temperature-ordered view of unresolved motions from the chromosphere to the transition region. 

These attributes, including layer selectivity, high contrast in cool-star spectra, rapid time response and a rich menu of line ratios, turn UV lines into \textit{bona fide} diagnostics of stellar activity.

\subsection{How should we interpret key diagnostic lines tracing stellar activity?}

\begin{figure}[t!]
\centering
\includegraphics[width=\columnwidth]{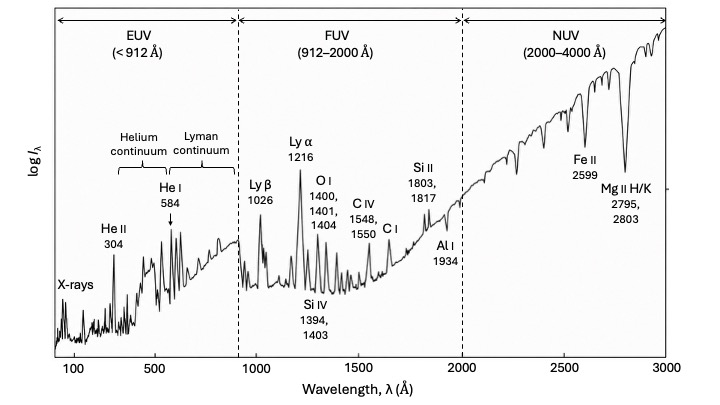}
\caption{Schematic rendition of the UV solar spectrum. EUV, FUV, NUV: Extreme, far- and near-UV. (After Rob Rutten's lecture notes; http://www.staff.science.uu.nl/\textasciitilde rutte101/rrweb/rjr-edu/lectures/rutten\_ssx\_lec.pdf.)}
\label{fig:solspec}
\end{figure}

We now provide a practical summary list of the key UV diagnostic lines. Refer to Figure \ref{fig:solspec} for a schematic overview of a selection of these lines.

\begin{itemize}

\item \textbf{Mg\,{\sc ii} $h$ (2803.53 {\AA})} 
  -- \emph{Region:} upper photosphere $\to$ chromosphere; 
  \emph{Blends/notes:} Fe\,{\sc i/ii} nearby, PRD; 
  \emph{Caveats:} Mg\,{\sc ii} $h/k$ deviates from a 2:1 ratio owing to opacity/PRD.
  
\item \textbf{Mg\,{\sc ii} $k$ (2796.35 {\AA})} 
  -- \emph{Region:} upper photosphere $\to$ chromosphere; 
  \emph{Blends/notes:} Fe\,{\sc i/ii} nearby, PRD; 
  \emph{Caveats:} strong opacity, core reversal (when the centre of an emission line is less intense than its wings, producing a dip at the line core), possible ISM absorption.

\item \textbf{C\,{\sc iii} (1906.68, 1908.73 {\AA})} 
  -- \emph{Region:} transition region; 
  \emph{Blends/notes:} Fe\,{\sc iii}/Si\,{\sc iii}; 
  \emph{Caveats:} requires near-UV coverage, higher extinction, useful for $n_\mathrm{e}$ (electron density) diagnostics.

\item \textbf{He\,{\sc ii} (1640.40 {\AA})} 
  -- \emph{Region:} photoionisation/recombination; 
  \emph{Blends/notes:} Fe\,{\sc ii} multiplets; 
  \emph{Caveats:} traces extreme-UV/X-rays, not a simple temperature diagnostic.
  
\item \textbf{C\,{\sc iv} (1548.20, 1550.77 {\AA})} 
  -- \emph{Region:} mid-transition region; 
  \emph{Blends/notes:} Fe\,{\sc ii}/Si\,{\sc i} in cool stars; 
  \emph{Caveats:} strong non-thermal wings.

\item \textbf{O\,{\sc iv} (1399.77, 1401.16, 1404.78 {\AA})} 
  -- \emph{Region:} transition region; 
  \emph{Blends/notes:} the line at 1404.78~{\AA} is blended with Si\,{\sc iv}; 
  \emph{Caveats:} de-blend carefully, useful for $n_\mathrm{e}$ diagnostics.

\item \textbf{Si\,{\sc iv} (1393.76, 1402.77 {\AA})} 
  -- \emph{Region:} mid-transition region; 
  \emph{Blends/notes:} inter-system blends, narrow ISM lines; 
  \emph{Caveats:} doublet ratio $<2$ indicates opacity/scattering.
  
\item \textbf{C\,{\sc ii} (1334.53, 1335.71 {\AA})} 
  -- \emph{Region:} lower transition region; 
  \emph{Blends/notes:} H$_2$/ISM blend. 
  \emph{Caveats:} often optically thick in active stars.

\item \textbf{N\,{\sc v} (1238.82, 1242.80 {\AA})} 
  -- \emph{Region:} hot transition region; 
  \emph{Blends/notes:} near Ly$\alpha$ wings;  
  \emph{Caveats:} faint in old/inactive stars.

\end{itemize}

This compact list contains the most frequently used UV diagnostics and their associated caveats. In practice, interpretation benefits from separating
\\(i) regions where a line forms; 
\\(ii) how it forms (optically thick non-LTE with scattering versus optically thin collisional/radiative); 
\\(iii) which plasma property it primarily constrains (heating, density, kinematics). 

Below, we outline a mechanism–agnostic reading of the diagnostics.

\noindent
1. \textit{Chromospheric resonance lines (Mg\,{\sc ii} $h$ and $k$).}

\noindent
Mg\,{\sc ii} $h$ and $k$ form across the upper photosphere and chromosphere in strong resonance scattering with PRD. The wings sample the deeper, warmer atmospheric layers and reflect the local temperature gradient and microturbulence; the line cores form in the higher, scattering-dominated layers ($\epsilon\!\ll\!1$). They often show central reversals caused by overlying cooler layers and flows \citep{Leenaarts2013a, Leenaarts2013b}. Core asymmetries (red or blue) of Mg\,{\sc ii} $h$ and $k$ are good tracers of bulk motions, whereas time variability encodes impulsive heating and subsequent relaxation. Because the opacity is high, profiles must be modelled with non-LTE radiative transfer.

\noindent
2. \textit{Transition-region resonance doublets (C\,{\sc iv}, Si\,{\sc iv}, N\,{\sc v}).}

\noindent
These lines are often well-described as optically thin collisional lines, with radiative emitters in plasma near their formation temperatures ($\log T \mbox{ [K]}\!\approx\!4.9$--5.3). Their integrated fluxes scale with the emission measure at $\sim\!10^{5}$\,K \citep{2019ApJS..241...22D,2021ApJ...909...38D}, thus enabling constraints on both heating and the differential (temperature-dependent) emission measure. They are prime tracers of chromospheric evaporation and turbulent broadening; line widths generally grow with increasing formation temperature. In very active stars, doublet ratios $<$2 reveal opacity and scattering.

\noindent
3. \textit{He\,{\sc ii}\,1640 {\AA} as an extreme-UV proxy.}

\noindent
He\,{\sc ii}\,1640 {\AA} is mainly powered by photoionisation/recombination and, thus, responds to the unobserved extreme-UV/X-ray radiation field as well as to transition-region conditions. Its strength, combined with a resonance-line tracer (e.g., C\,{\sc iv}), can be used to reconstruct the stellar extreme-UV budget relevant to atmospheric escape and photochemistry in exoplanetary systems \citep{France2018}. Because it is not a single-temperature diagnostic, it should be interpreted within a multi-line context.

In summary, stellar activity encompasses the suite of processes that add non-radiative energy and momentum to the outer atmosphere. It builds chromospheres, transition regions and coronae and drives in- and outflows. In cool stars, it is chiefly magnetic; in hot stars, it originates from radiation forces, pulsation and accretion. Observationally, stellar activity appears in the form of excess line emission, profile asymmetries/broadening and variability across a wide range of timescales. UV lines trace this activity, because their strongest transitions form where the energy is actually deposited---that is, between the upper chromosphere and the transition region---under low densities that make formation highly non-local and sensitive to small perturbations. The associated high line-to-continuum contrast and advances in photon-counting time resolution turn modest changes in heating and flows into useful diagnostic spectroscopic signals.

%%%%%%%%%%%%%%%%%%%%%%%%%%%%%%%%%%%%%%%%%%%%%%%%%%%%%%%%
\section{UV spectroscopy of pre-main-sequence stars}
\label{sec3}

% Ana I. Gomez de Castro: variability of PMS stars

PMS stars are significantly brighter in the UV than their main-sequence descendants, largely owing to their larger radii, ongoing gravitational contraction and enhanced activity (accretion and shocks). At early evolutionary stages, their UV luminosities can reach up to $0.2 L_{\odot}$, particularly when accretion and bipolar outflows begin to disperse the surrounding cocoon of gas and dust. The UV diagnostics of PMS stars depend strongly on stellar mass and evolutionary stage. In particular, the spectral tracers observed in low-mass T Tauri stars (TTSs; $M \leq 2 M_{\odot}$) differ from those characteristic of the more massive Herbig Ae/Be (HAeBe) stars. The main spectroscopic differences from their main-sequence analogues arise from the contribution by powerful outflows and accretion discs.

\subsection{Low-mass PMS stars} 

TTSs become detectable at UV wavelengths typically at ages of $\sim 1$~Myr. Their strong UV emission is primarily driven by accretion \citep{2013ApJ...767..112I,2016ARA&A..54..135H}, with gravitational energy being released in to form of shock fronts at the stellar surface. Although the UV luminosity declines as the star approaches the main sequence, even at ages of $\sim 100$ Myr, low-mass PMS stars still display spectral features characteristic of highly active stars, similar to those of the M-type emission-line (dMe/Me) stars.

\begin{figure}[t!]
\centering
\includegraphics[width=\columnwidth]{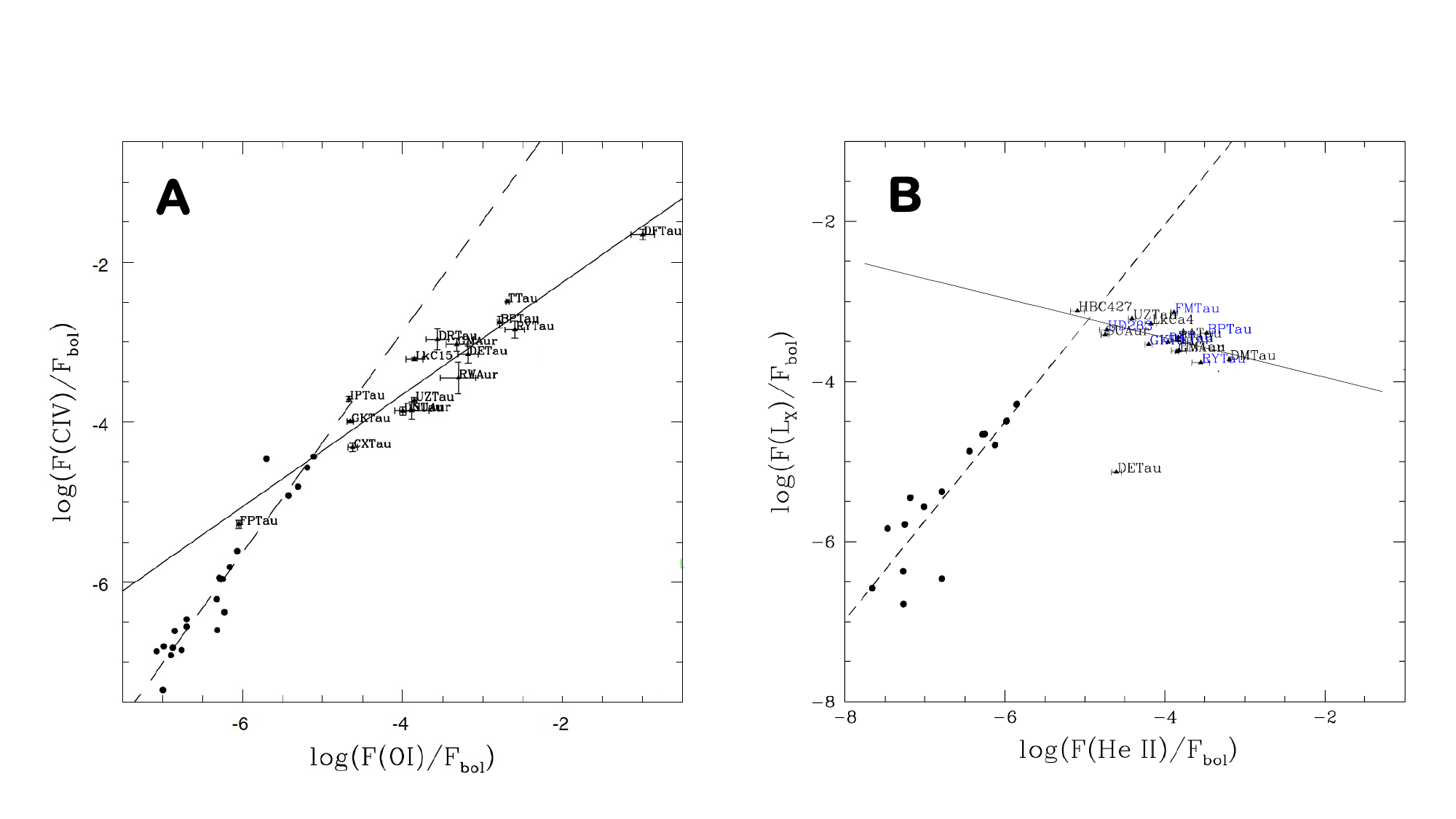}
\caption{(A) Comparison of the C\,{\sc iv} to O\,{\sc i} flux ratios of TTSs and main-sequence stars. (B) Comparison of the X-ray to He\,{\sc ii} flux ratios of TTSs and main-sequence stars. The main-sequence UV fluxes are from \citet{1995ApJS...96..223A} for G0 to K4 types and from \citet{1982ApJ...260..670L} for K5 to M5.6 types. \citep[Credit:][]{2012ApJ...749..190G} }
\label{fig:civoi}
\end{figure}

The UV spectra of low-mass PMS stars share spectral diagnostics with magnetically active stars (see Section \ref{sec2}). The surface fluxes of these lines are typically $10^{6-7}$ erg~cm$^{-2}$~s$^{-1}$, approximately 2 orders of magnitude higher than those observed in the Sun \citep{1980ApJ...239L.115I,1992A&A...257..671L, 1997iueu.book.....G}. The distinct characteristics of PMS stars become especially evident when compared with their main-sequence counterparts. As illustrated in the standard flux--flux diagrams (see, e.g., Figure~\ref{fig:civoi}) used to study energy transport in the atmospheres of cool stars, there is notable excess emission of singly ionised species which is not caused by metallicity effects (all TTSs are members of the Taurus--Auriga star-forming region). High-resolution spectroscopic studies have measured a mean iron abundance close to solar, [Fe/H] $\simeq 0.0$ dex, with only a small star-to-star dispersion. This suggests that the Taurus population is chemically homogeneous and does not contain a significant metal-rich component \citep{2011A&A...526A.103D}. This emission originates from an extended plasma region surrounding the star, which is not directly heated by collisional processes or by photoionisation from the stellar surface. Part of this excess is attributed to unresolved outflows, whereas another portion is thought to arise from the star's extended magnetosphere---an area that can span several stellar radii and channels accretion flows onto the stellar surface. In this sense, a notable feature is the distinct relationship between X-ray and He {\sc ii} emission and the relatively high X-ray luminosity in PMS stars compared with their main-sequence counterparts, reflecting differences in magnetic activity, accretion processes and energy dissipation mechanisms during the PMS phase.

Flux--flux relations are powerful diagnostics for disentangling the relative contributions of accretion and other heating processes. This was demonstrated through the monitoring of the PMS binary AK Scorpii (AK Sco), a close binary system surrounded by a circumbinary disc, where the strongest accretion events occur near periastron. Observations obtained during two consecutive periastron passages follow distinct tracks in the flux--flux diagrams, a behaviour interpreted as evidence that the relative importance of accretion with respect to other heating mechanisms varies from epoch to epoch \citep{2020ApJ...904..120G}.

\begin{figure}[t!]
\centering
\begin{tabular}{cc}
\includegraphics[width=0.47\columnwidth]{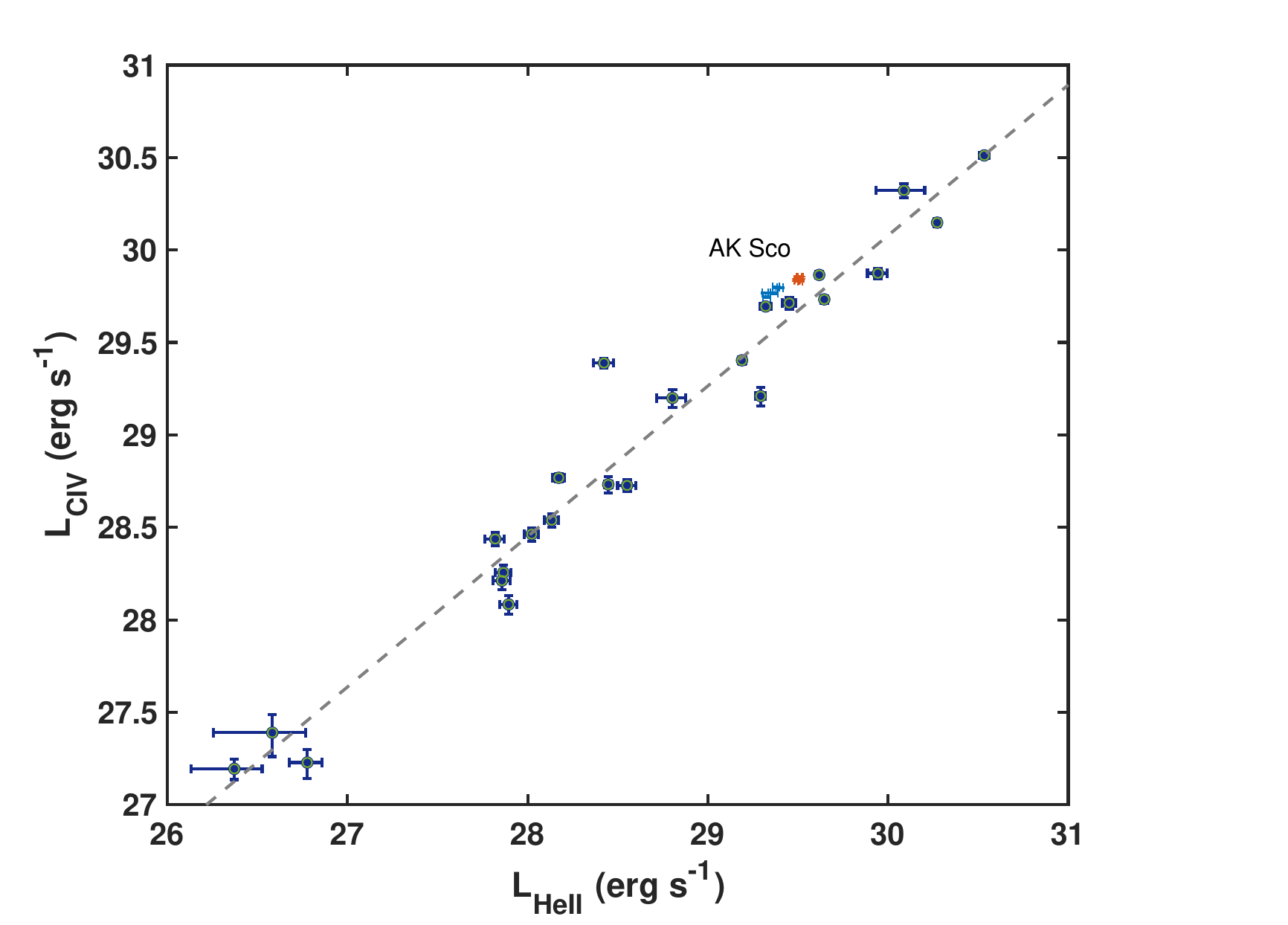} & \includegraphics[width=0.45\columnwidth]{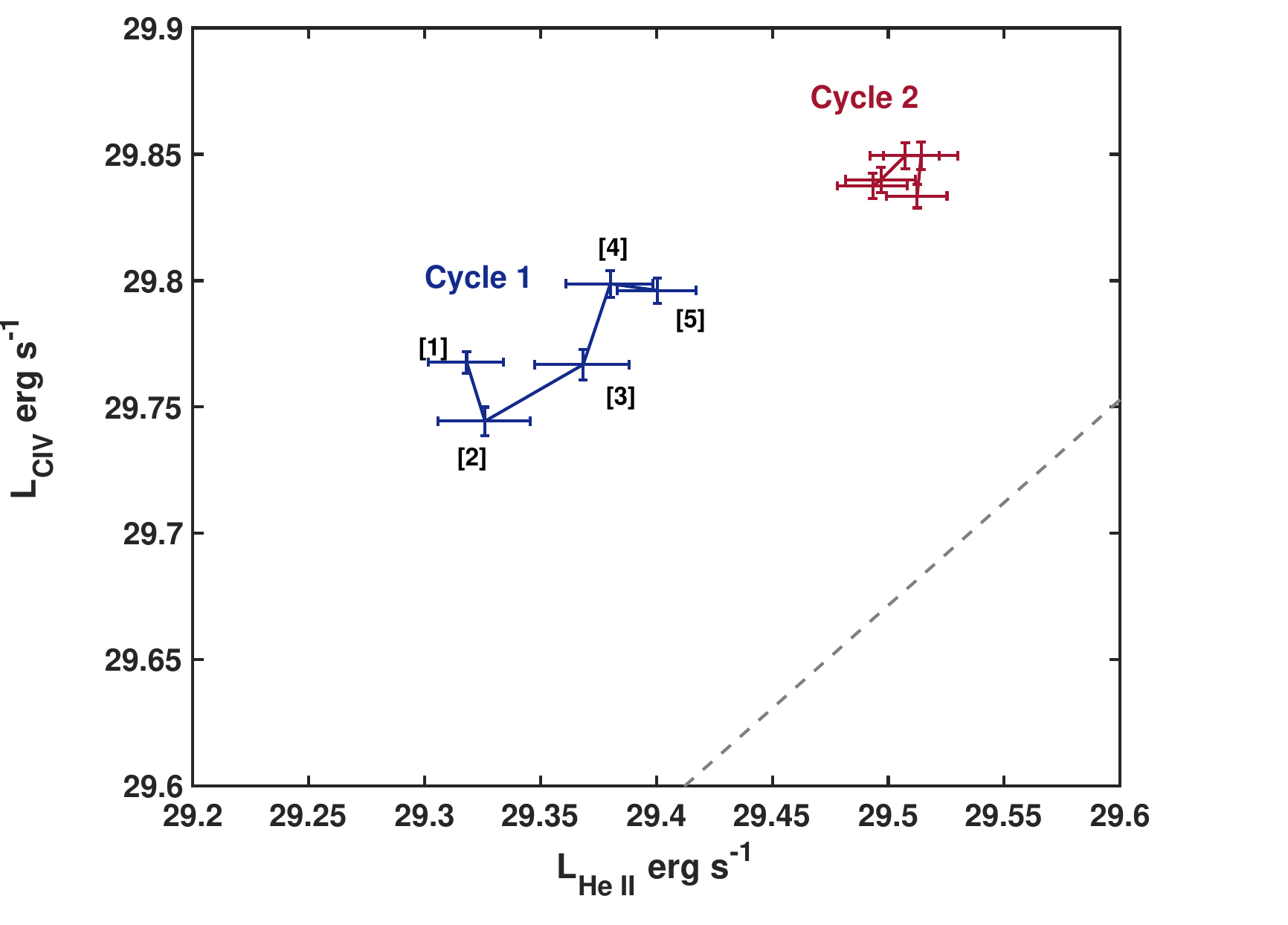}\\
\end{tabular}
\caption{C\,{\sc iv}--He\,{\sc ii} flux--flux diagrams. (left) All TTSs were observed with the {\sl HST}, using the same configuration (STIS/G140L). AK Sco observations are shown in blue (Cycle 1) and red (Cycle 2). The best-fitting regression line is also included. (right) Enlargement of the location of AK Sco observations. The order of the observations in Cycle 1 is marked by [1], [2], ...; 1$\sigma$ error bars are shown \citep{2020ApJ...904..120G}.}
\label{fig:aksco_ff}
\end{figure}

Another significant spectroscopic characteristic is the occurrence of prominent Fe\,{\sc ii} emission. Many multiplets are observed, including UV 2, 3, 35, 36 (2330--2410 {\AA)}; UV 32, 62, 63 (2700--2750~{\AA}); and UV 60, 78 (2900--3000 {\AA}) \citep{1979A&A....73L...4G, 1984MNRAS.207..831B, 1997iueu.book.....G}. Although the Fe\,{\sc ii} lines are individually weaker than the equivalent Mg\,{\sc ii}, C\,{\sc ii} or Si\,{\sc ii} lines, they are so numerous that their combination becomes a significant coolant of the PMS stellar atmospheres \citep[see, e.g.,][]{1988ESASP.281b..97J}. The excitation mechanism of the transitions is not purely collisional; recombination following photoionisation is also an important pumping route that is especially significant for transitions pumped by Ly$\alpha$ photons---such as the Fe\,{\sc ii} transition at 2507 {\AA}. The latter account for as much as 80 per cent of the total far-UV flux of TTSs.

The spectral lines display significant supra-thermal broadening which decreases as the stars evolve towards the main sequence \citep{2013ApJS..207....1A, 2013ApJ...775..131G}. This broadening is often interpreted as comprising two components, including a broad component associated with magnetospheric accretion and a narrow component likely originating from the heated stellar atmosphere. As the TTSs approach the main sequence, the broadening decreases until it reaches the thermal broadening observed in main-sequence stars

Intercombination (semi-forbidden) transitions are also observed in many spectra. In some cases they are produced in unresolved jets \citep{2001ApJ...548..976G,2014MNRAS.442.2951L}, but they mostly originate from extended magnetospheric structures located between the star and the inner boundary of the accretion disc---such as the ion torus detected in RW Aurigae (RW Aur) \citep{2003ApJ...597..443G}. The C\,{\sc iii}, Si\,{\sc iii}] and O\,{\sc iii}] transitions have also been used as diagnostic tools to probe accretion shocks \citep{1999MNRAS.304L..41G}. The strongest transitions are:

\begin{itemize}
\item C\,{\sc ii}] 2324.21, 2325.40, 2326.11, 2327.64 and 2328.84 {\AA} [2s$^2$2p ($^2$P$^0$) $\to$ 2s2p$^2$ ($^4$P)];
\item Si\,{\sc ii}] 2328.52 and 2334.41 {\AA} [3s$^2$3p ($^2$P$^0$) $\to$ 3s3p$^2$ ($^4$P)];
\item C\,{\sc iii}] 1908.73 {\AA} [2s$^2$ ($^1$S) $\to$ 2s2p ($^3$P$^0$)];
\item Si\,{\sc iii}] 1892.03 {\AA} [3s$^2$ ($^1$S) $\to$ 3s3p ($^3$P$^0$)]; and
\item O\,{\sc iii}] 1660.81 and 1666.15 {\AA} [2s$^2$2p$^2$ ($^3$P) $\to$ 2s2p$^3$ ($^5$S$^0$)].
\end{itemize}

\begin{figure}[t!]
\centering
\includegraphics[width=10cm]{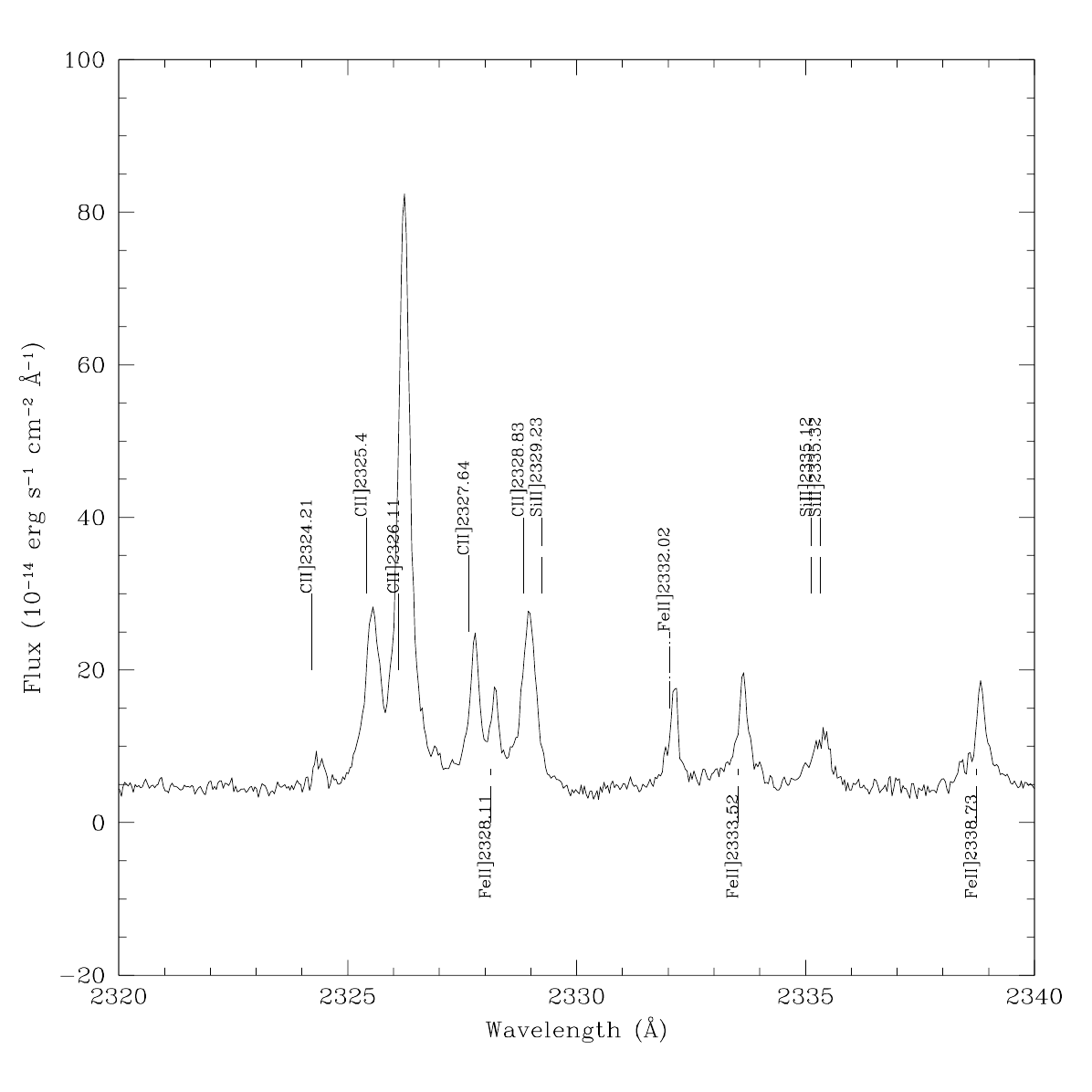}
\caption{The C\,{\sc ii}], Fe\,{\sc ii}] and Si\,{\sc ii}] blend observed in TW Hya. The line broadening in this spectrum is thermal, and the individual features are well-resolved \citep{2014MNRAS.442.2951L}. }
\label{fig:c2_twhya}
\end{figure}

Figure~\ref{fig:c2_twhya} shows example C\,{\sc ii}] and Si\,{\sc ii}] profiles; the contribution of Fe\,{\sc ii}] transitions is also detected [3d$^6$4s (5D) $\to$ 3d64p(5D)]. From careful analysis of such profiles, important information about the prevailing magnetospheric plasma properties in the magnetospheric flow can be measured, such as the electron temperatures  ($10^{4.1-4.5}$ K) and the range of electron densities, spanning from $10^8$~cm$^{-3}$ to $10^{12}$~cm$^{-3}$. The line widths are supra-thermal, and the broadening correlates with the accretion rate; only two TTSs displayed thermal broadenings, including CY Tauri (CY Tau) and, more interestingly, TW Hydrae (TW Hya; shown in Figure~\ref{fig:c2_twhya}). In a few stars the line centroids are blue-shifted, indicating that the line emission is dominated by outflows. 

The molecular gas in circumstellar discs also produces notable signatures in UV spectra \citep{2002ApJ...572..310H,2004ApJ...607..369H,2012ApJ...756..171F, 2013ApJ...766...12M}. The most abundant molecules, H$_2$ and CO, exhibit strong electronic transitions that are pumped by Ly$\alpha$ photons. In particular, the transitions from the ground state ($X^1\Sigma_\mathrm{g}^+$) to low-lying excited states---Lyman ($2\mathrm{p}\sigma,B^1\Sigma_\mathrm{u}^+$) \citep{1993A&AS..101..273A} or Werner ($2\mathrm{p}\pi,C^1\Pi_\mathrm{u}$) \citep{1993A&AS..101..323A} levels, i.e. the two primary electronic excited-state systems of molecular hydrogen, H$_2$, that produce the characteristic far-UV absorption and emission bands. The CO transitions are also pumped by Ly$\alpha$ photons by means of several rotational transitions of the $A^1\Pi$ $\to$ $X^1\Sigma^+$ (fourth positive) electronic system.

Beyond the emission-line diagnostics discussed above, the UV continuum itself provides a valuable probe of PMS evolution. Classical TTSs frequently exhibit a pronounced UV excess relative to the photospheric continuum expected for a main-sequence star of the same spectral type. This excess is generally attributed to hot accretion shocks formed where magnetically channelled material impacts the stellar surface, producing continuum emission that can dominate the near-UV spectrum. Measurements of this UV excess are widely used to estimate accretion luminosities and mass-accretion rates. In contrast, weak-lined TTSs typically show continua much closer to photospheric expectations, with chromospheric activity contributing primarily through line emission rather than a strong continuum excess \citep{1998ApJ...509..802C, 1998ApJ...492..323G, 2013ApJ...767..112I}. The UV continuum of TTSs generally declines with age as accretion rates decrease, evolving from the strong accretion-driven UV excesses observed in classical TTSs to the weaker, predominantly chromospheric emission characteristic of weak-line TTSs (see G\'omez de Castro et al., this volume).

\subsection{Herbig Ae, Be stars}

HAeBe stars are intermediate-mass (2--$10 M_{\odot}$) PMS stars, whose formative processes and outer atmospheres remain the subject of ongoing research. Although their relatively high masses suggest that their gravitational collapse should be super-Alfv\'enic---thus rendering magnetic fields a minor driver---accumulated observations imply a more nuanced view. Their UV emission-line spectra indicate the presence of a chromosphere above the photosphere \citep{2005A&A...429..247D}, and the detection of highly ionised species (transition-region or corona-like), together with the detection of weak magnetic fields \citep{1997MNRAS.291..658D,2019MNRAS.489..886J}, imply that magnetic drivers are indeed active during at least the first $\sim 5 \times 10^6$~yr of PMS evolution. They remain rather puzzling objects. 

Monitoring campaigns in the UV and optical ranges first revealed azimuthal structures in the wind of AB Aurigae \citep[AB Aur;][]{1986ApJ...303..311P}. Subsequent extensive spectroscopic monitoring \citep[e.g., the {\sc musicos} campaigns;][]{1998cvsw.conf..349C} revealed that chromospheric structures rotate with a period of 32 hours (consistent with the stellar rotation period), whereas the wind (traced by UV Mg\,{\sc ii} lines) exhibits a longer, 45 hour period \citep{1996A&AS..120..431B,1999A&A...345..884C}. UV studies have also identified clumps of very hot gas in the AB Aur wind, traced by N\,{\sc v} emission \citep{1997A&A...328..606B}. These features strongly suggest a multi-component wind structure, which can be described as follows:
\begin{itemize}
\item A `slow' dense outflow, attaining terminal velocities of  $\sim 300$~km~s$^{-1}$, responsible for the prominent P Cygni profiles in Ca\,{\sc ii} and Mg\,{\sc ii} uv1 lines (the strongest resonance transitions of a singly ionised metal in the UV) and the broad blue-shifted absorption seen in C\,{\sc iv}. Semi-empirical modelling allows inference of mass-loss rates of order $10^{-8} M_{\odot}$~yr$^{-1}$ 
\citep{1987A&A...174..158C, 1998A&A...340..163B}.
\item A `high-velocity' component, composed of magnetically confined streamers forming a faster outflow. Because HAeBe stars are rapid rotators, gas in the streamers is forced into co-rotation out to the Alfv\'en radius, the distance where the stellar-wind speed equals the Alfv\'en speed. This leads to shocks between the slow and fast wind components and the formation of dense azimuthal structures at co-rotating interaction regions.
\end{itemize}

Magnetically confined flows are further supported by detections of Ly$\alpha$ emission from the jets associated with the HAeBe stars HD 163296 and HD 104237 \citep{2000ApJ...542L.115D,2004ApJ...608..809G}. Moreover, magnetic dissipation provides a plausible mechanism for the radiative losses observed in the chromosphere and wind of about 4–-8 per cent of the stellar bolometric luminosity \citep{1998A&A...340..163B}, although accretion streams may also contribute \citep{1993A&A...268..624B,2000A&A...359.1011B, 2001ApJ...551L..97R}. Although accretion likely does not drive the bulk of the outflow, radiatively driven winds can become collimated in the presence of a modest magnetic field \citep{1985A&A...152..121S,1996ApJ...464..859R}. The origin of the magnetic field remains an open question; possibilities include a dynamo operating in the outer stellar layers, with turbulence induced by pulsations---radial and non-radial, with periods from tens of minutes to hours \citep[see][]{2003Ap&SS.284...53C}---or by shear generated through rotational wind braking \citep{1996A&A...314..465L}.

Recent UV spectroscopic investigations have provided deeper insights. High-resolution UV spectra of a small HAeBe sample show that the luminosity of the C\,{\sc iv} doublet scales (albeit with scatter) with the estimated accretion rate, thus extending trends found for lower-mass stars. Nevertheless, the line shapes are consistent with optically thin winds rather than magnetospheric accretion flows \citep{2016ApJ...825..147C}. Mid-infrared forbidden-line studies (e.g., [Ne\,{\sc ii}] and [Ne\,{\sc iii}]) also suggest that high-energy irradiation (UV, extreme-UV, X-rays) plays a key role in shaping the circumstellar environments of HAeBe stars \citep{2025MNRAS.540.3330A}.

Combined, these results support a paradigm where HAeBe stars, despite their higher masses and more rapid evolution, host structured chromospheres and magnetically mediated winds that bear both similarities to and distinctions from the lower-mass T Tauri class. The extant evidence emphasises that magnetic fields, although perhaps weaker or more complex, remain important during the early phases of intermediate-mass PMS evolution; that hot plasma and structured winds are significant; and that UV emission lines remain a powerful diagnostic of the outer stellar environment and its dynamics. Future progress will benefit from high-resolution UV spectroscopy and coordinated multi-wavelength campaigns in the intermediate-mass regime.

\section{Low-mass stars in the UV: Observational status} 
\label{sec4}

% Deepak Chahal; covering the photosphere, chromosphere and corona of FGKM stars

Cool stars with convective envelopes---particularly those of spectral type F and later---exhibit magnetic activity throughout their atmospheres. In the Sun, where we can resolve surface features, activity features manifest as large, rapidly evolving starspots and faculae (bright regions) in the photosphere, prominences and filaments in the chromosphere, and energetic flares or coronal mass ejections (CMEs) from the corona \citep[e.g.,][]{Berdyugina2005,Brun2015,deGrijs2021}. Multiwavelength diagnostics, including photometric variability tracing photospheric activity, chromospheric emission lines such as Ca\,{\sc ii} H and K, H$\alpha$ and Mg\,{\sc ii}, as well as UV/X-ray flux measurements from the corona, provide a comprehensive framework for probing the generation and evolution of these phenomena. Low-mass stars display varying levels of magnetic activity over their lifetimes, thus making it essential to understand how evolutionary processes shape these changes. Key open questions include how stellar rotation spins down with time, how magnetic activity evolves with age and which fundamental parameters govern these relationships. To date, most studies have concentrated on studying individual activity indicators, often limited by observational constraints. However, the advent of large-scale time-domain surveys now offers unprecedented opportunities to investigate magnetic activity across multiple wavelengths, particularly in the optical and UV domains.

Optical and near-UV emission originates from the cool ($T\approx 10^{4}$ K), lower stellar atmosphere, whereas far-UV lines trace the hotter ($T\approx 10^{5}$ K) transition-region plasma \citep{Kowalski2024}. Owing to this sensitivity to hot plasma, the UV regime serves as a key diagnostic of stellar magnetic activity. The first extensive observations of chromospheric emission were obtained by {\sl IUE}. Essentially all solar-type stars showed the presence of a chromosphere (traced by the line emission of, e.g., the Mg\,{\sc ii} doublet at 2800\,{\AA}, the C\,{\sc iv} doublet at 1550\,{\AA}, Si\,{\sc iv} at 1394\,{\AA} and many other, mostly weaker, lines). \citet{Giampapa1981} reported strong emission features in cool dwarfs, whereas\citet{Hartman1979} and \citet{Haisch1980}  identified intense flare-driven enhancements in stars such as Proxima Centauri ($\alpha$ Centauri C/$\alpha$ Cen C) and EQ Pegasi (EQ Peg). \citet{1981MNRAS.197..815B} further showed that active M dwarfs (e.g., Gl 867 A and AU Microscopii/AU Mic) exhibit prominent UV emission lines, indicative of magnetically heated atmospheres. 

{\sl GALEX} enabled population-wide studies of stellar UV emission in both near-UV (1771--2831 {\AA}) and far-UV (1344--1768 {\AA}) bands (note that these definitions of the near- and far-UV are {\sl GALEX}-specific), covering key diagnostic lines such as C\,{\sc ii}, C\,{\sc iv}, Mg\,{\sc ii} and Fe\,{\sc ii}. {\sl GALEX} also opened the time-domain frontier, revealing variability in M-dwarf flares \citep{2007ApJS..173..673W} and RR Lyrae pulsations \citep{2005ApJ...619L.123W}. More recently, \citet{Bai2018} studied optical activity from the Large Sky Area Multi-Object Fiber Spectroscopic Telescope (LAMOST), combined with UV behaviour from {\sl GALEX}, providing a statistical bridge between optical and UV diagnostics tracing magnetic activity.

UV flux measurements have long been central to understanding stellar activity and its influence on planetary atmospheres. UV radiation drives key photochemical reactions, including ozone formation and molecular dissociation, thus shaping planetary habitability \citep{Miguel2014,Harman2015}. Whereas excessive near-UV flux can destroy biomolecules and O$_{3}$ \citep{Rugheimer2015}, moderate UV levels enable prebiotic chemistry \citep{Rimmer2018,Ranjan2020}. Recent studies have revealed that even optically inactive cool stars exhibit strong UV variability, often linked to flares \citep{France2018,Loyd2018,Jackman2023}, yet long-term UV monitoring remains limited. Future dedicated UV missions are essential to trace stellar activity cycles, age--activity relations and the evolving radiation environments of exoplanet-hosting stars.

\subsection{Active Variables Catalogue}
In the past decade, photometric time-series surveys from space ({\sl CoRoT}, {\sl Kepler}, {\sl K2}, {\sl TESS}) and ground-based telescopes (OGLE, the Optical Gravitational Lensing Experiment; ASAS-SN, the All-Sky Automated Survey for SuperNovae; ZTF, the Zwicky Transient Facilty) have provided an extensive catalogue of variable stars. Most low-mass stars exhibit starspot/faculae modulation in their light curves, which is quasi-periodic in nature owing to the redistribution/evolution of spots and faculae. The availability of long-term, high-cadence observations for millions of stars has enabled the detection of rotational modulation in large samples of low-mass stars, as summarised in Table \ref{table:RotVarCat}. These catalogues present a unique opportunity to exploit multiwavelength analysis (from the photosphere to the corona) to constrain the evolution of stellar activity across the low-mass stellar regime. However, comparatively fewer time-series observations have been obtained in the UV (e.g., {\sl GALEX}) and in X-rays ({\sl eROSITA}, {\sl Chandra}, {\sl NICER}). Upcoming UV missions (see Section \ref{modernUV} and Barstow et al., this volume) will enable comprehensive optical--UV time-domain studies of these rotational variables, allowing detailed characterisation of the flare properties, activity cycles and related phenomena discussed below. These observations will, in turn, provide statistical constraints on the generation and evolution of stellar magnetic activity in low-mass stars.

% \usepackage{booktabs}
\begin{table}[t!]
\caption{Catalogue of rotational variables detected from space-borne and ground-based time-series observations}
\label{table:RotVarCat}
\centering
\begin{tabular}{@{}cc@{}}
\toprule
\multicolumn{2}{c}{Space-borne} \\ \midrule
{\sl TESS} \citep{Fetherolf2023} & 68,497 \\ %\midrule
{\sl Kepler} \citep{Nielsen2013,McQuillan2014,Reinhold2023} & 12,253; 34,030; 20,397 \\ %\midrule
{\sl K2} \citep{Armstrong2015,Reinhold2020} & 9399; 32,387 \\ %\midrule
{\sl Gaia} \citep{Distefano2023} & 474,026 \\ \midrule
\multicolumn{2}{c}{Ground-based} \\ \midrule
OGLE \citep{Iwanek2024} & 18,443 \\
ZTF \citep{Chen2020,Chahal2022,Lu2022} & 81,393; 40,553 \\
ASAS-SN \citep{Jayasinghe2020} & 81,393 \\
\bottomrule
\end{tabular}
\end{table} 

\subsubsection{Flares from the optical to the UV} 

Stellar flares are impulsive releases of magnetic energy that produce broadband photon emission from gamma-rays to radio wavelengths, with UV and optical bands exhibiting strong line emission and continuum fluxes that jointly dominate the radiative energy budget \citep{Osten2005,Fuhrmeister2008}. Observations from the HAZMAT programme \citep{2014AJ....148...64S} and the MUSCLES survey \citep{France2016,2017ApJ...843...31Y} revealed that the quiescent (basal) far-UV/near-UV luminosity increases by nearly 2 orders of magnitude from early-K to late-M dwarfs, and even optically inactive M dwarfs display frequent UV flares, as inferred from H${\alpha}$ and Ca\,{\sc ii} variability \citep{France2013,Loyd2018,Diamond-Lowe2021,Feinstein2022}. \citet{Brasseur2023} identified numerous short-duration near-UV flares in {\sl GALEX} and {\sl Kepler} data \citep{2005ApJ...619L...1M,Borucki2010}. Most lacked optical counterparts, thus demonstrating that rapid UV-bright flares are often missed in low-cadence optical observations. Coordinated high-cadence optical monitoring combined with rapid-response UV observations (e.g., with {\sl HST} or {\sl Swift}) is therefore essential to capture the impulsive flare phases and construct contemporaneous optical--UV light curves. The large catalogue of rotational variables listed in Table \ref{table:RotVarCat} offers an opportunity to investigate flare properties across a broad range of stellar masses and ages, emphasising the need for multiwavelength, high-cadence monitoring---particularly from the optical to the UV---to fully characterise stellar flare energetics. A more detailed discussion on stellar flares is presented in Section \ref{sec5}.

\subsubsection{Stellar Activity Cycles}

Stellar magnetic activity often exhibits cyclic variations driven by the underlying dynamo processes, exemplified by the 11-year solar cycle \citep{Baliunas1995,Dikpati2006,Charbonneau2010,deGrijs2021}. Such cyclic behaviour has been studied extensively across a broad range of low-mass stars \citep{Wilson1968,1978ApJ...226..379W,1984ApJ...279..763N,Baliunas1995,Brandenburg1998,Saar1999,Olah2016,Reinhold2017,BoroSaikia2018,Olspert2018,Isaacson2024,Chahal2025}. Nevertheless, the mechanisms governing the generation and evolution of stellar activity cycles remain poorly constrained, even for the Sun and Sun-like stars. Activity cycles are expected to depend sensitively on stellar rotation, mass, metallicity and other fundamental stellar parameters. Multiwavelength observations are essential to probe these cycles in detail. For instance, \citet{Wargelin2017} studied $\alpha$ Cen C’s activity cycles across optical, UV and X-ray bands. A major challenge in detecting solar-like cycles is the limited time coverage of existing data sets. Recent work by \citet{Chahal2025} combined photometry from {\sl Kepler} Full-Frame Images \citep{Montet2017}, ASAS-SN and ZTF, thereby enabling the detection of longer-term cycles. Additionally, ASAS-SN alone \citep{Irving2023} also provides a valuable long-baseline data set. These extensive photometric surveys allow the construction of a large sample of rotationally modulated activity cycles (see Table \ref{table:RotVarCat}), which can be followed up in the UV and at X-rays to probe multiwavelength activity tracers. Such studies offer critical insights into solar and stellar dynamos and can inform our understanding of habitable environments around other stars.

\subsection{Correlation of Activity with Rossby number}
\label{rossby.sec}

The stellar dynamo responsible for magnetic activity can be indirectly probed through the correlation between coronal emission and stellar rotation. This relation is commonly expressed in terms of the X-ray luminosity ($L_\mathrm{X}$) or the UV flux as a function of the Rossby number and is inherently model-dependent. The X-ray activity--Rossby number relation exhibits two regimes, a saturated regime for rapid rotators and an unsaturated regime for more slowly rotating stars. Early X-ray and chromospheric studies already demonstrated the tight connection between rotation, Rossby number, and magnetic activity \citep[e.g.,][]{1981ApJ...248..279P, 1984ApJ...279..763N, 1984A&A...133..117V, 1994A&A...292..191S}. \citet{Pizzolato2003} later formalised this framework using a large, homogeneous {\sl ROSAT} sample, clearly delineating the saturated and unsaturated regimes, and subsequent work \citep[e.g.,][]{Wright2011,Stelzer2013,Stelzer2016,Wright2016,Astudillo-Defru2017,Houdebine2017,Newton2017,Shulyak2017,Wright2018,Magaudda2020} has further refined the empirical picture.

\begin{figure}[t!]
\centering
    \includegraphics[width=0.48\linewidth]{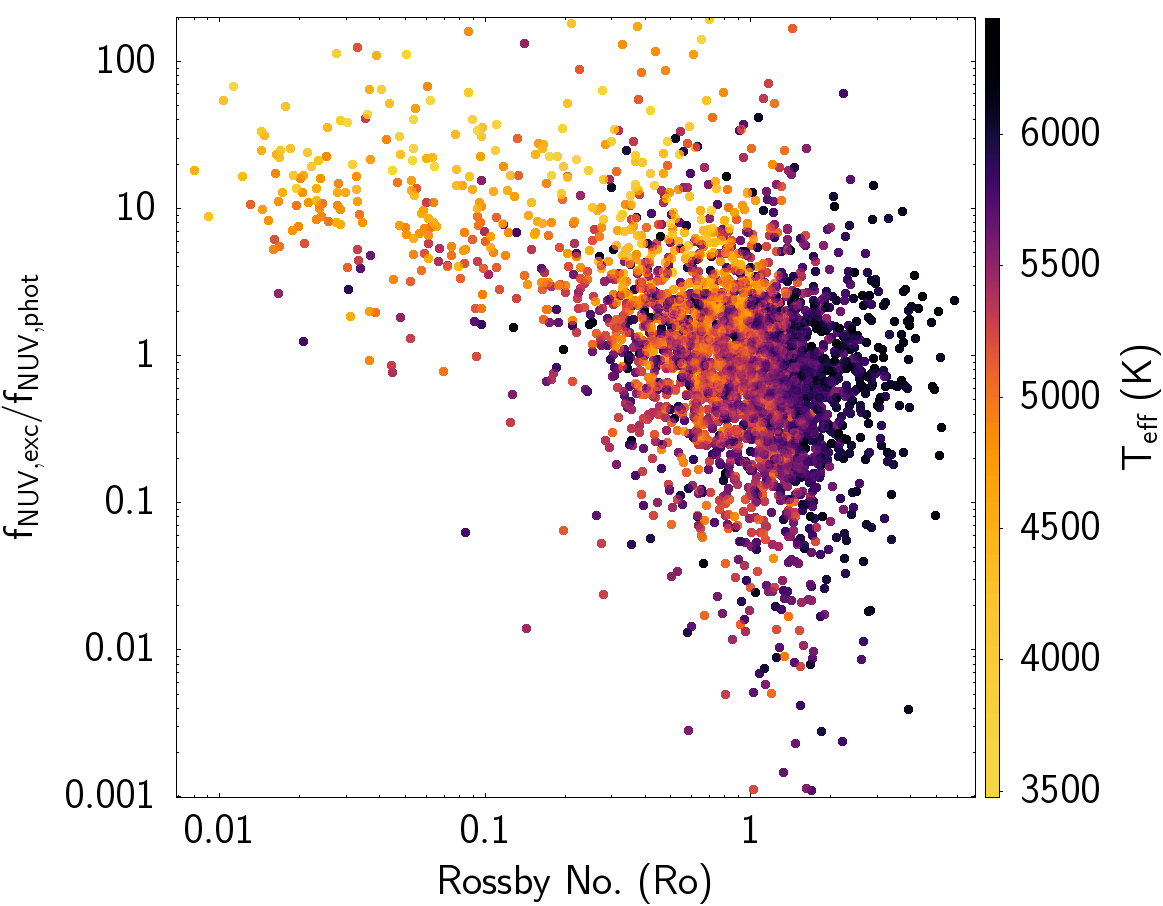}
    \includegraphics[width=0.48\linewidth]{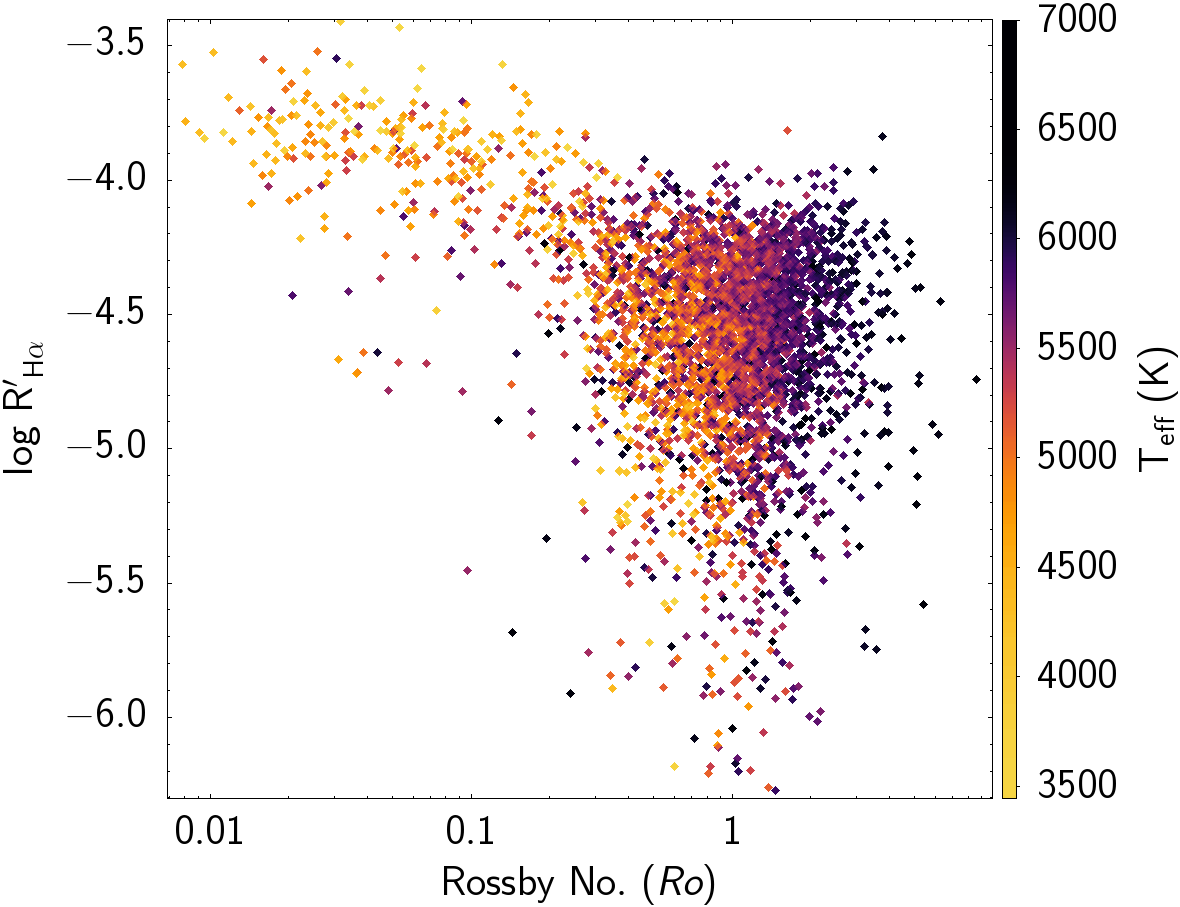} 
    \caption{Rossby number ($\equiv P_{\mathrm{rot}}/\tau_{\mathrm{conv}}$) plotted against (left) the fractional UV excess and (right) H$\alpha$ activity indices. UV-excess values were derived from {\sl GALEX} photometry \citep[after][]{Bai2018,Li2024}, and the H$\alpha$ indices were measured from LAMOST spectra \citep[both from][]{Chahal2023}. The colour markers represent the stellar effective temperature, as indicated by the colour scale.}
    \label{fig:4}
\end{figure}

While correlations between stellar activity and rotation have been studied extensively in X-ray and optical bands, relatively few authors have examined the corresponding trends in the UV regime \citep[e.g.,][]{Vardavas2005,Pineda2021}. The Far Ultraviolet M-dwarf Evolution Survey (FUMES) \citep{Pineda2021,Youngblood2021,Duvvuri2023,Duvvuri2025} recently advanced this field by investigating UV emission, flare properties and their dependence on rotation for M dwarfs. Building on this progress, \citet{Chahal2023} analysed the chromospheric activity of $\sim$16,000 low-mass stars as a function of rotation period and Rossby number (see Figure \ref{fig:4}, right). Figure \ref{fig:4} presents the fractional near-UV flux excess (left) derived from {\sl GALEX} data \citep[following][]{Bai2018,Li2024} and the H$\alpha$-based chromospheric activity (right) as functions of the Rossby number, where $\tau_{\mathrm{conv}}$ was estimated using the empirical relation from \citet{Cranmer2011}. Both indicators exhibit a saturated regime at low Rossby numbers, transitioning to an unsaturated regime as the Rossby number increases; however, the latter does not display the same linear decline in activity typically observed in the X-ray activity--Rossby number diagram. These relations can be further refined through improved estimates of Rossby numbers and more precise measurements of UV and chromospheric activity indices derived from high-resolution spectra. Stellar metallicity is also expected to play a significant role in the generation of stellar activity \citep{Karoff2018}; better constraints on metallicity can help reduce the scatter in these relations. Exploring such multiwavelength correlations provides valuable insights into the structure and evolution of stellar atmospheres and their underlying magnetic activity processes.

\section{Stellar flares}
\label{sec5}
%% Fabio Favata

\subsection{Solar-type stars}

`Solar-type stars' are stars with a structure similar to our Sun, i.e. with a convective envelope and a radiative core. They encompass, roughly, the main sequence from F to M stellar types.\footnote{Note that the coolest M dwarfs do not have a radiative core; they are fully convective.} The key phenomenology that grants their definition as a class is the presence of an outer, hot atmospheric layer above their photosphere. The photosphere is optically thick and heated directly by the heat flux from the stellar interior. Pure thermodynamics arguments would imply that the temperature should decrease above the photosphere. Yet, in our Sun we clearly see---when the Sun is observed at UV, extreme-UV and X-ray wavelengths---that the temperature above the photosphere increases, first gradually to tens of thousands of K (in the chromosphere) and then sharply to hundreds of thousands of K (in the transition region), reaching millions of K in the extended outer layer, the corona.

The driving force that creates this hot extended atmosphere is the Sun's magnetic field, generated by our star's internal dynamo, which is likely driven by the interaction between the Sun's (differential) rotation and convection in the outer layer. The Sun's magnetic field is relatively weak in the photosphere, where gas pressure dominates magnetic pressure (i.e. $\beta \gg 1$), so that the photosphere is almost entirely structured by gravity and stellar rotation (even though sunspots are a visible effect of the magnetic field's concentration). 

As we ascend from the photosphere, and thus encounter hotter plasma, the gas remains optically thin, and it increasingly departs from spherical symmetry. In fact, its pressure decreases (owing to the prevailing, much lower ambient density), so that one transitions from a regime in which $\beta \gg 1$ to a regime in which $\beta \ll 1$, i.e. one in which the plasma is structured and confined by the magnetic field. The magnetically confined plasma is constantly heated by magnetic-field reconnection (through processes whose details are still not entirely clear), resulting in an outer atmospheric regime that is much hotter than the photosphere.

The Sun is thus far from being the perfect, unchanging object of Aristotelian physics, and thanks to these mechanisms it emits across a range of wavelengths much more energy than would be implied by its relatively modest photospheric temperature. These wavelengths range from the UV all the way to energetic gamma-rays, as well as non-thermal radio emission. This broad non-photospheric emission, the `stellar activity', is powered by the combination of convection and rotation.

X-ray and extreme-UV observations from space have shown that the Sun's magnetic fields confine the plasma into thin `loops', semicircular structures emerging from sunspots and creating `plasma bottles' within which the plasma moves essentially uni-dimensionally, along the field lines. The first outer atmospheric layer is the chromosphere, immediately above the photosphere, which, while optically thin (unlike the photosphere) reaches about 20,000--35,000\,K some 2000\,km above the photosphere. Above the chromosphere lies the transition region, where the temperature rises sharply to some 1--5\,MK some 100--500\,km above the chromosphere. The transition region is followed by the corona, which in the Sun hosts plasma at temperature of several MK. Both the transition region and the corona are optically thin.

Whereas the chromosphere is still gravity-structured, so that it can be envisaged as a spherical layer overlying the photosphere, the corona is highly structured, with the plasma concentrated in magnetically confined loops. And whereas coronal plasma typically emits in X-rays (given its temperature), chromospheric and transition-region plasmas emit in the UV to extreme-UV regime. These structures can be observed and studied in detail in the Sun thanks to the very high spatial resolution at a range of wavelengths and specific plasma diagnostics reached by the current generation of solar space telescopes.

\begin{figure}[t!]
\centering
\includegraphics[width=\textwidth]{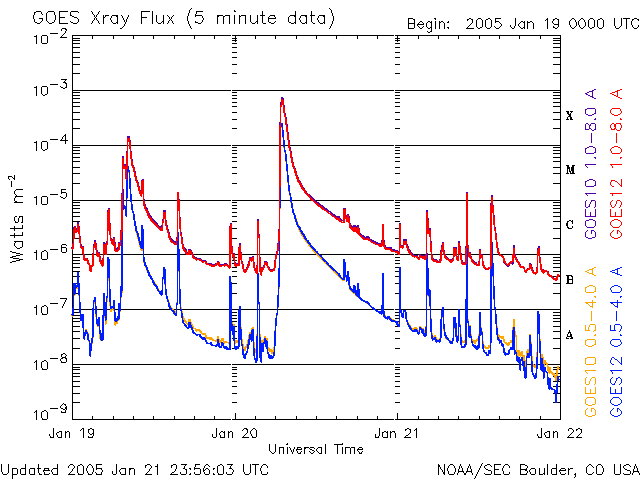}
\caption{Example {\sl GOES} solar X-ray flux (5-min averaged data) recorded between 19 and 22 January 2005, showing solar soft X-ray emission in two wavelength bands: 0.5--4.0 {\AA} (yellow and blue) and 1.0--8.0 {\AA} (purple and red). Several major solar flares are evident as sharp flux increases, with classification levels (A, B, C, M, X: very weak, small, moderate, large and major flares, respectively) indicated on the right-hand axis. (Data from the U.S. National Oceanic and Atmospheric Administration/Space Environment Center {\sl GOES} 10 and {\sl GOES} 12 satellites.)}
\label{goes-data}
\end{figure}

Although the Sun's photospheric emission is extremely constant, with a modulation of parts in 10,000 during the solar cycle owing to the varying area covered by sunspots and faculae, the emission caused by stellar activity is highly variable, with the degree of variability increasing with decreasing wavelength (and thus increasing plasma temperature). In fact, the solar X-ray emission, which has been monitored continuously for decades by the {\sl GOES} (Geostationary Operational Environmental Satellite) series of space-weather satellites, hardly shows a basal level of emission, and visual inspection of the light curves shows them to be the result of the superposition of a continuous series of impulsive events of varying peak luminosity and duration: for a relevant example, see Figure \ref{goes-data}.

These impulsive events show a typical fast, almost impulsive rise (pointing to an impulsive heating event) followed by a slower, exponential decay. They are broadly grouped under the general category of `flares'. Detailed space-borne observations of the Sun which commenced with the Solar Maximum Mission ({\sl SMM}) have enabled detailed studies and understanding of the flaring mechanism. It typically starts with the release of impulsive energy close to the top of the flaring structure, resulting in a flow of particles that travels almost freely along the magnetic field lines in the corona, until it hits the denser chromosphere and photosphere. At that point, the lower layers are heated and plasma evaporates, filling the flaring loop with hot plasma, which will radiate in X-rays (while cooling) through its radiation as well as its conduction to the lower atmosphere.

Flares in the Sun are observed to have a power-law intensity distribution \citep[e.g.,][]{2022Physi...5...11S}, with a large number of small flares and a small number of rarer, more intense flares. Solar flares have been observed across a wide range of wavelengths, from radio waves to visible light, UV, X-rays and gamma-rays. Spatially resolved observations of solar flaring events made possible by space observatories, pioneered by {\sl SMM} in the early 1980s, enabled the development of a `standard model' of solar flares. In this model \citep[see, e.g.,][their Figure 13]{2017LRSP...14....2B}, the plasma in a coronal loop is heated impulsively at the top by a magnetic reconnection event, with the energy release consisting of accelerating particles. The energetic particles then precipitate to the chromosphere, where they heat the plasma to the high temperatures observed in soft X-rays. The hot plasma then expands along the loop into the corona, a process referred to as evaporation.

When energetic electrons (and possibly ions) hit the dense underlying chromosphere via Coulomb collisions (interactions between charged particles mediated by the electrostatic force, causing energy and momentum exchange), they lose energy, the temperature in the chromosphere increases and the resulting pressure exceeds the ambient chromospheric pressure, with plasma thus flowing back into the loop. The evaporated hot plasma appears to be the source of most of the observed soft X-ray emission. This qualitative scenario explains the features observed in most (but not all) solar flares and has served remarkably well as explanatory framework for decades.

\subsection{Lower-mass stars}

Stellar activity, including flaring, is common in solar-type stars, with the relevant phenomena becoming proportionally more intense in lower-mass stars. Traditionally, a specific class of stars was defined, flare stars or UV Ceti-type variable stars, and the first instance of significant variability in the optical range was reported by Adriaan van Maanen (1884--1946) in 1945 \citep[see the discussion in][]{1954PASP...66....5J}. This was based on photographic-plate observations of WX~Ursae Majoris (WX UMa) and YZ~Canis Minoris (YZ CMi), which were observed to brighten by about 1.5 mag for some tens of minutes. Once photoelectric detectors allowed continuous stellar light-curve monitoring, flaring appeared to be common in low-mass stars, with the flaring intensity increasing with decreasing wavelength, i.e. with flares manifesting more strongly in the blue and UV ranges.

In fact, modern high-cadence observations of low-mass stars show that their near-UV emission, both in the continuum and the lines (e.g., the Ca\,{\sc ii} H and K lines) is the superposition of a near-continuous series of flaring events of varying intensity, as for example shown by \citet{2016ApJ...820...95K} for YZ~CMi and by \citet[][their Figure 1]{2024LRSP...21....1K} for $\alpha$ Cen C. Once one moves to the UV regime, flaring is therefore no longer an exceptional event but rather the normal regime where these stars emit in the UV (as well as in higher-energy bands).

In the UV the flaring emission observed is the result of two different phenomena, with very different characteristics. On the one hand, there is the optically thin emission from the heated, optically thin chromosphere that is typically observed in the form of line emission, with a few lines dominating the UV spectrum between Ly$\alpha$ and the visible range. On the other hand, the photosphere, being dense and optically thick, will form a hotspot where the energy has been deposited, which will radiate the energy thermally, exhibiting a (near-)blackbody spectrum. Emission in the extreme-UV is produced largely within the transition region, with mechanisms similar to those operating in the chromosphere.

The Ly$\alpha$ line (at 1216\,{\AA}) is by far the strongest emission line in the solar UV spectrum, and in low-mass M dwarfs it carries by itself as much energy as the rest of the UV spectrum combined \citep{2012ApJ...750L..32F}. However, its use as a diagnostic is limited, by both astrophysical challenges (because of its strength, the assumption of optical thinness is easily challenged) and observational challenges from low-Earth orbit, where geocoronal emission from the Earth's atmosphere often contaminates the observed Ly$\alpha$ flux.

Stellar flares in the UV generally exhibit rapid time evolution \citep[e.g.,][]{2019ApJ...883...88B}. However, one should keep in mind that the number of stellar flares observed with good time coverage and high-quality data is limited, given that flares are stochastic events and their observation is largely a matter of luck. In fact, only a very small number of flares have been observed in the UV, and even fewer have been analysed through their rise and decay phases \citep[for a list of the observations, see the review by][]{2024LRSP...21....1K}, so that the detailed behaviour of the Sun's chromospheric emission during intense flares is still not known in sufficient detail.

The behaviour of the photosphere is, by contrast, much better known thanks, first, to the numerous ground-based campaigns to observe flaring stars in the $U$ band and also to the large number of stars that have been monitored in white light while searching for exoplanetary transits by {\sl Kepler} and {\sl TESS}. Ground-based campaigns have been carried out with relatively small telescopes, so that long observing campaigns could be dedicated to individual objects, thus significantly raising the likelihood of observing a large, well-sampled flaring event. These observations essentially monitor in a limited number of bands ranging from the atmospheric cut-off in the UV to the red part of the spectrum. This can effectively be combined with observations of the emission in the Ca\,{\sc ii} lines and in the Balmer series \citep[for a well-studied flare on the M dwarf YZ~CMi, see, e.g.,][]{2013ApJS..207...15K}, which thanks to their being ground-based can reach a coverage and data quality thus far very rare for space-borne observations.

Although the detection of stellar flares in the {\sl TESS} and {\sl Kepler} light curves is done in white (visible) light, these detections are relevant in the context of UV astronomy because they cover the red tail of the blackbody emission from the heated stellar photosphere. The blackbody temperature will be such that the bulk of its radiation is emitted in the UV. The temperature of the photospheric blackbody can only be inferred through broad-band colourimetry (which often includes the Balmer jump), and values ranging from $T \approx 8500$\,K up to 50,000\,K have been reported in the literature. At such temperatures, the bulk of the emission will be in the UV, even though this is, because of purely observational constraints, rarely observed.

Despite concentrating on the tip of the proverbial iceberg (white-light emission), {\sl TESS} and {\sl Kepler} observations have enabled the study of statistically large samples of stars, resulting in a number of studies of, e.g., the frequency distribution of flares as a function of their energy output, as well as the high-energy tail of the distribution, where rare, large-scale events take place. Some optical flares detected by {\sl TESS} and {\sl Kepler} show peak enhancements of the white-light output of the star of tens of per cent of the star's quiescent luminosity. Given that only a small fraction of the flare emission will be emitted in white light (depending on the photospheric blackbody temperature), we can only imagine how glaringly bright such events will be when observed in the UV and hope that an upcoming generation of space-borne observatories will enable UV observations of such events.

In fact, whether or not the Sun’s photospheric emission can be adequately described by a blackbody spectrum remains fiercely debated. Modern high-precision irradiance measurements reveal systematic deviations caused by line blanketing and radiative-transfer effects \citep[e.g.,][]{2003SoPh..214....1T,2005oasp.book.....G,2018A&A...611A...1M}. Although increasingly sophisticated models have been developed, incomplete wavelength coverage and calibration uncertainties continue to limit progress \citep{2016A&A...588A.126C}.

In conclusion, while decades of observations, both ground-based and space-borne, have advanced our knowledge and understanding of stellar flaring events quite significantly, very few large flares have been observed in detail in the UV. This is important not only because of our understanding of stellar atmospheres, but also because many phenomena such as atmospheric erosion (also caused by CMEs), photochemistry, etc. are driven by the activity of the parent stars. Understanding the dynamics, and the emission, of the most energetic events is thus necessary to understand the evolution of exoplanets, with implications for questions pertaining to habitability. We thus hope that a new generation of cost-effective space-borne UV observatories will enable the observation and understanding of stellar flares to advance significantly.

\section{UV studies of Supernovae: An early-time perspective}
\label{sec6}
%% Firoza Sutaria; progenitor models and UV flashes

\subsection{Shock breakout}

Massive stars, with masses $> 8.5 M_{\odot}$, end their exothermic lives through gravitational core collapse. This eventually results in an SN explosion, leaving behind remnants that are both compact (e.g., a neutron star/NS, a quark star or even a stellar mass black hole/BH) and nebular (the SN remnant). The SN shock transports the gravitational binding energy of the collapsing stellar core ($GM^2/r \sim 10^{51}$ erg) from the edge of the proto-NS to the outer layers of the progenitor. In the process, it imparts both thermal and kinetic energy to the outer layers of the erstwhile progenitor, ionising them and pushing them outwards. For matter at a local optical depth $\tau$ and a physical depth $d$, radiation trapping occurs as long as the diffusion timescale of the photons ($\propto \tau^2 d/c$) is shorter than the shock-propagation timescale, and radiation pressure contributes to shock propagation. As the shock plows into layers with lower optical depth, photons from the heated ejecta diffuse out faster than the local shock velocity, $v$. At this point, a radiative precursor exhibits a sharp surge in luminosity since thermal X-ray and UV photons break out of the shock front. This situation, which occurs when $\tau \lesssim c/v$, is known as `shock breakout' and marks the beginning of the SN explosion \citep[see][and references therein]{Klein+Chevalier-1978,Blinnikov_et_al-2013,2017hsn..book..967W}. For $v$ in the range $10^{3-4}$ km\,s$^{-1}$, this surge occurs at $10 \lesssim \tau \lesssim 100$, i.e. well below the stellar photosphere at $\tau=1$. Following SBO, the stellar envelope expands and cools nearly adiabatically, resulting in shock-cooling emission. 

The fireworks that result from this terminal, explosive event are thus heralded by a flash (of $\sim 10^{3-4}$ s duration) of photons whose non-Planckian spectrum \citep{Falk-1978} peaks in the soft X-ray band, although with significant contributions in the far- and near-UV bands. In most stars with extended outer layers, the diffusion timescale ($t_\mathrm{diff}$) is comparable to the dynamical timescale ($t_\mathrm{dyn}$, defined as the time required for the break-out layer to expand to twice its thickness), and thus the diffusion region remains tied to the Lagrangian coordinate (i.e. the mass element) where the SBO occurred. The physical radius at the SBO, the duration of the flash and the peak luminosity of the burst will all depend on the total energy radiated, which is itself a function of shock strength as well as of the progenitor's density and opacity profile. In turn, they will also determine the characteristics of the resulting light curves and spectra.

Note that SBO is expected even in thermonuclear (Type-Ia) SNe, which occur via runaway thermonuclear burning in low-mass ($< 8 M_{\odot}$), degenerate (He or CO white-dwarf) progenitors, members of either singly or doubly degenerate binary systems. The explosion model for these events is still under debate, with both deflagration \citep[see][and references therein]{Fink_et_al-2014} and detonation \citep[see][and references therein]{Seitenzahl_et_al-2013, Pollin_et_al-2025} mechanisms considered plausible. However, the detection of SBOs in Type-Ia SNe would also play an important role in reconciling the two possible mechanisms, via the delayed detonation transition model, which would begin as a subsonic deflagration and transition into a supersonic detonation \citep[see][and references therein]{Piro_et_al-2010}. Initial simulations by \citet{Piro_et_al-2010} show that the SBO in this brightest class of SNe would be extremely fast ($\sim 10^{-2}$ s), faint ($\sim 10^{40}$ erg) and peak in the hard X-ray ($\sim 20$ keV) band. Alternatively, if the progenitor's structure allows $t_\mathrm{dyn} \gg t_\mathrm{diff}$, planar cooling could increase the overall radiated luminosity, thus making its detection possible. Such a situation could occur during SBO in Type-Ia SNe with CO progenitors, initiated via a detonation of the outer He layer, followed by a final, terminal detonation of the CO layer \citep{Piro+Nakar-2025}. This double detonation transition mechanism results in an aspherical SBO, with a relatively faint X-ray surge of $\lesssim 10^{41}$ erg lasting only $\lesssim 10$ s. However, this is followed by soft X-ray-bright ($10^{43}$ erg), short-lived ($\sim 4$ s) post-break-out planar cooling, unique to progenitors with He outer layers, and then 12--24 hr-long UV-dominated shock-cooling emission.

Core-collapse SNe (CC-SNe) comprise a zoo of distinct subtypes, resulting from stripped-envelope progenitors (Types Ib/Ic) as well as progenitors with extended H envelopes (Types II-P/L). Additionally, massive stars often undergo mass loss during their post-main-sequence evolution via line-driven winds or convective overshooting (when convective motions penetrate into adjacent stable layers beyond the formal convective boundary, causing extra mixing and affecting stellar structure and surface phenomena), resulting in the formation of a circumstellar medium (CSM) whose density profile and morphology traces its evolutionary history. Narrow-line SNe result from these progenitors, cocooned within an extended, dense CSM, and they are highly diverse in their presumed pre- and known post-explosive properties \citep[to explore CC-SNe diversity, see][for Type-IIn]{Hiramatsu_et_al-2024}; for Type-Icn and for CC-SNe in general, see \citet{Pellegrino_2022} and \citet{Maeda_et_al-2012}, respectively. Finally, nearly $\sim$50 per cent of massive progenitors are expected to have had a binary companion at some point during their evolution \citep{Zapartas_et_al-2019}. As can be expected, the SBO characteristics of CC-SNe are also diverse. Several studies have focussed on SBOs in these progenitors, the implications of which are briefly discussed below.

\subsection{UV Modelling at the Supernova Shock}

Theoretically, both analytical and numerical attempts have been made to estimate the strength and duration of UV emission at the SBO for CC-SNe. Three-dimensional (3D) simulations of SBO in red-giant progenitors of Type-IIP SNe \citep{Goldberg_et_al-2022}, the most numerous sub-class of CC-SNe, show an asymmetric breakout duration of $\sim$3--6 hr, spanning a range of radiative temperatures. This is in agreement with observations \citep[e.g.,][]{Gezari-2008, Gezari-2015}, although the predicted luminosities are somewhat lower ($L_\mathrm{bol} \sim 10^{44}$ erg s$^{-1}$). Their 3D simulations appear to eliminate the possibility of using observed rise times to measure the stellar radius via light-travel time effects. The effect of a dense CSM on the SBO was investigated by \citet{Balberg+Loeb-2011}, who used a forward--reverse shock structure for shocked ejecta interacting with the pre-existing wind to establish an analytical relation between the SBO radius and velocity, which they used to estimate the progenitor radius of the X-ray/UV-bright Type-Ibc SN2008D. Their results are in agreement with direct estimates from early X-ray and UV detections of this event \citep{Rabinak+Waxman_2011}. 

More detailed numerical simulations have been carried out by \citet{DeLaRosa-2017}, who assumed a single-temperature (for matter and radiation) photon diffusion model coupled with an homologous outflow (the {\sc lanl} code) to model the UV light curves of the Type-IIn SNe (with well-studied precursor activity) SN 2009ip and SN 2011ht from SBO to the late nebular phase. They used progenitor masses in the range of 15--35 $M_{\odot}$ for SN2009ip and 20--45 $M_{\odot}$ for SN2011ht and found that while the peak of the UV light curve is determined by the shock temperate and the explosion energy, $E_\mathrm{expl}$, the profile depends on the mass of the ejecta as well as on $E_\mathrm{expl}$. 

Next, a simulation of the number of events from which SBOs would be detected with a dedicated X-ray+UV wide-field survey instrument has been carried out by \citet{Bayless_et_all-2022}. They modelled the UV emission from CC-SNe in red-supergiant (RSG), blue-supergiant (BSG) and a range of Wolf--Rayet (WR) progenitors using the same code, assuming solar metallicity and incorporating electron scattering-dominated `grey' opacity. They found that the duration of the UV flash at the SBO ranges from only 400 s for some RSG models to up to 10,000 s for some WR calculations (see their Table 2). Additionally, assuming a limiting UV flux of $8.0 \times 10^{-14}$ erg s$^{-1}$ cm$^{-2}$, they determined that over a 2 yr period, $\sim 5000$ events with RSG progenitors, $\sim 3700$ events with WR progenitors but only $\sim 200$ events with BSG progenitors could be detected at the 3$\sigma$ confidence limit (see their Table 4).   

We now address the question of the actual detection of the SBO's UV component. In the absence of X-ray/UV surveys of the transient sky (although the {\sl Einstein Probe} X-ray satellite and the proposed {\sl ULTRASAT} mission look promising in this context), direct UV detection of SBOs has been mainly serendipitous. Although early-time UV detections of several Type-II events are available---notably, Type-IIpec SN 1987A, Type-IIb SN 1993J and Type IIP SN 2006bp, as well as multiple events in the {\sl Swift}/UVOT bands---these have all been detected in the shock-cooling phase. The earliest detections of actual SBO (i.e. UV brightening) were obtained for the Type-IIP SNe SNLS-04D2dc and SNLS-06D1jd, discovered by the Supernova Legacy Survey \citep[SNLS; see][and references therein]{Gezari-2008}, followed by the {\sl GALEX} near-UV detection of the Type-IIP SN PS1-13arp \citep{Gezari-2015}. Shock--CSM interaction has been detected since then in both early and late phases. Notably, the Type-IIn SN2007pk was detected by {\sl Swift}/UVOT $\sim 5$ days post-SBO \cite{Pritchard_2012}. Note, however, that SBOs in interacting SNe (of which Type-IIn and Type-IIbn are subtypes) depend on the poorly understood density profile of the transition region between the progenitor and the CSM. \cite{Inkenhaag_etal-2025} searched the fields of view of 51 CC-SNe from their {\sl HST} snapshot survey, taken within five years of discovery. They found late-time (2--4 yr) UV rebrightening only for ASAS-SN-17qp and ATLAS17lsn, which they attribute to very late-time interaction of the SN shock with the CSM. 

The very early-time spectra of SNe across all classes appear to be dominated by a continuum which peaks in the UV. In many instances, also superimposed are narrow lines emitted from overlying, slowly moving material, often tenuous CSM. The narrow features may be fleeting (i.e. appearing for only a few days or, at most, a week) for all but the class of `interacting' SNe, the best-known of which are the Type-IIn. SBO in Type-IIn SNe---whose  progenitors have extended CSM---is expected to be delayed owing to their extended winds, but, assuming a density profile for the wind, can yet be used to constrain the combination of explosion energy and ejecta mass \citep{Balberg+Loeb-2011}. Whereas the flash ionisation lines themselves are indicators of the morphology and composition of the CSM, however tenuous, this is probably the only epoch in a SN's lifetime when a thermal continuum can be unequivocally identified and isolated in its spectra, thus permitting direct estimation of the velocity and temperature at the SBO. By extension, assuming reasonable values for $E_\mathrm{expl}$, these can be correlated with progenitor radius and mass. 

In summary, at the start of the emerging era of large-scale time-domain surveys (including the Legacy Survey of Space and Time undertaken with the Vera Rubin Observatory), when rapid follow-up of transients will soon become the norm, there is an imperative need for dedicated rapid follow-up, highly sensitive, X-ray and UV transient `factories' to make full use of this opportunity to study the final moments of the poorly understood final stages of the evolution of massive (and less massive) stars.

\section{Beyond the Milky Way: tracing star formation in the Magellanic Clouds using UV diagnostics}
\label{sec7}

Our Galaxy is accompanied by many dwarf galaxies, including the Large and Small Magellanic Clouds (LMC and SMC, respectively). Located at distances of $D_\mathrm{LMC}\sim$50 kpc and $D_\mathrm{SMC}\sim$62 kpc \citep{2014..de..Grijs,2015..de..Grijs,2019..Distance...LMC,2020ApJ...904...13Graczyk..SMC..dist}, their close proximity enables observation of their resolved stellar populations. Moreover, their metal-poor environments \citep[$Z_\mathrm{LMC} = 0.5\, \mathrm{Z}_\odot$ and $Z_\mathrm{SMC} = 0.2\, \mathrm{Z}_\odot$;][]{2014..Dobbie..Z..SMC,2018..Choudhury..SMC,2021..Choudhury..Z..LMC,2023..Hocd..Z..LMC..SMC..MW} provide conditions analogous to those in the early Universe. Together, these properties render the Magellanic Clouds among the best laboratories for investigating how low-metallicity environments influence star formation \citep{1998..Kennicutt,2024..Hota..SMC}, massive-star evolution and the stellar feedback processes that regulate galaxy evolution.

Their young stellar populations encode the galaxies’ recent evolutionary history, preserving signatures of tidal interactions, morphological transformations and the physical mechanisms regulating star formation. The ongoing dynamical interaction of the Magellanic Clouds with each other and with the Milky Way \citep{2009..Harris..SFH..MCs,2018..Rubele..MCs..int,2022...SD..Massana} continues to shape their current star-forming activity, thus linking the study of the Magellanic Clouds not only to local galaxy evolution but also to processes that governed star formation in the early Universe. In addition, their rich populations of young massive stars provide a direct link between recent star formation and the progenitor populations of transient sources, including core-collapse SNe and other endpoints of massive-star evolution, including neutron stars and black holes. Consequently, UV studies of the Magellanic Clouds also offer key insights into the environments that shape stellar activity and the production of transient events \citep{2009ARA&A..47...63S,2013..Conroy}.

UV observations are particularly powerful tracers of recent star formation, since galactic emission from young, massive stars dominates the UV continuum \citep{1998..Kennicutt,2013..Conroy}. Unlike in optical or near-infrared bands, where evolved stellar populations contribute significantly, the UV is far less contaminated by older stars \citep{2007..Salim}. This sensitivity allows UV diagnostics to probe young stellar populations directly, offering insights into the intensity and spatial distribution of recent star formation. Moreover, because massive stars are the primary drivers of stellar feedback through ionising radiation, stellar winds and SN explosions, UV observations provide valuable insights into the processes that shape the surrounding ISM and regulate subsequent star formation. UV observations are especially valuable in low-metallicity systems such as the Magellanic Clouds, where the reduced dust content mitigates extinction compared with more metal-rich galaxies \citep{2003..Gordon}. These conditions also influence the evolution of massive stars, their feedback mechanisms and the nature of the compact remnants and transient phenomena they ultimately produce. In this section, we highlight how UV observations have advanced our understanding of the recent star-formation history, spatial structure, kinematics and hierarchical organisation of young stellar populations in the Magellanic Clouds.%, while providing insight into the low-metallicity environments that shape stellar activity and transient progenitors.} 

Over the course of the last three decades, consecutive UV space missions, from the Ultraviolet Imaging Telescope \citep[{\sl UIT};][]{1998..UIT.Parker,1997..Cornett..UIT} through {\sl GALEX}, {\sl Swift}/UVOT and the {\sl AstroSat}/Ultra Violet Imaging Telescope \citep[UVIT;][]{2017June..TondonUVIT,2017Sep..TondonUVIT}, including the {\sl HST}, have provided valuable insights into the recent star-formation activity in the Magellanic Clouds and the characteristics of their you ng massive stars. These surveys have enabled investigations ranging from individual massive stars and stellar clusters to galaxy-wide studies of recent star formation and stellar structure. The far-UV catalogue compiled by \citet{1998..UIT.Parker} and \citet{1997..Cornett..UIT} includes 37,333 sources in the LMC and 11,306 sources in the SMC. The near-UV catalogue of the LMC of \citet{2014..Simons} based on {\sl GALEX} data contains around 6 million sources. More recently, \citet{2025..Ludwig} published UV catalogues for both Magellanic Clouds using UVOT, covering a total of 734,869 objects. Together, these catalogues provide an unprecedented census of UV stellar populations across the Magellanic Clouds, enabling detailed studies of their recent star-formation history, stellar clustering, spatial structure and the properties of young massive stars.

Among current UV facilities, UVIT offers a major advance for the study of the Magellanic Clouds owing to its high spatial resolution of $1''.2$--$1''.4$ and simultaneous far- and near-UV imaging capabilities. The improved resolution relative to earlier UV surveys enables the identification and characterisation of individual young stellar populations and star-forming structures across the Magellanic System. Using UVIT observations, \citet{2024..Hota..SMC} identified approximately 76,800 sources across the SMC's main body and inner Wing. In a complementary study, \citet{2024..Hota..SMC..Shell} presented a far-UV catalogue of about 14,400 stars in the Shell region, located in the northeastern outskirts of the SMC. A summary of UV catalogues of the Magellanic Clouds is presented in Table~\ref{tab:UV_catalogs_MCs}.

\begin{table}[t!]
    \centering
    \caption{UV catalogues of the Magellanic Clouds.}
    \begin{tabular}{ccc}
    \hline
      Telescope& Catalogue & Comments\\
      \hline
       {\sl UIT}  & $N_\mathrm{LMC}=37,333$, $N_\mathrm{SMC}=11,306$ & Far-UV filter; spatial \\ & & resolution $\sim3''$ \\ & & \citep{1998..UIT.Parker,1997..Cornett..UIT} \\
       {\sl GALEX}  & $N_\mathrm{LMC}\approx6,000,000$ & Near-UV filter; spatial \\ & & resolution $\sim5''$ \\ & & \citep{2014..Simons}\\
        {\sl Swift}/UVOT & $N_\mathrm{LMC}=471,507$, $N_\mathrm{SMC}=263,362$ & Three UV filters; spatial \\ & & resolution $\sim2''$ \\ & & \citep{2025..Ludwig} \\
        {\sl AstroSat}/UVIT & $N_\mathrm{SMC}\approx76,800$, $N_\mathrm{SMC\,Shell}\approx14,400$ & Far-UV filter; spatial \\ & & resolution $\sim1''.4$ \\ & & \citep{2024..Hota..SMC..Shell,2024..Hota..SMC} \\
        \hline
    \end{tabular}
    \label{tab:UV_catalogs_MCs}
\end{table}

\subsection{Recent Star Formation History, Morphology, and Kinematics of Young Stellar Populations in the Magellanic Clouds}

Early UV observations revealed that recent star formation in the Magellanic Clouds is highly structured and occurs across a hierarchy of environments, including OB associations, stellar clusters and stellar complexes. Analysing four {\sl UIT} fields covering the SMC bar, \citet{1997..Cornett..UIT} identified four concentrations of UV-bright stars arranged from northeast to southwest with nearly uniform separations of about 0.5 kpc. By comparing the large-scale far-UV and H$\alpha$ morphologies, they inferred a spatial and temporal progression of star formation, unfolding over the past $\sim$20 Myr. Similarly, \citet{1998..UIT.Parker} reported that the {\sl UIT} fields encompass 47 OB associations in the LMC and 37 in the SMC, along with 102 and 74 H\,{\sc ii} regions from the catalogue of \citet{1976MmRAS..81...89D}. The far-UV data allow easier identification of the hottest OB stars compared with optical photometry, and their ionising flux shows a clear correlation with H$\alpha$ emission. They found continuous star formation up to $\sim$500 Myr for the field stars, while the star cluster populations are characterised by a best-fitting age of around 3.4 Myr, consistent with young OB associations drawn from multiple clusters. \citet{1992..Parker..N11} obtained photometry for more than 1200 stars and spectra for 43 O and 30 early-B stars in the OB associations LH 9 and LH 10. LH 10 is younger, more reddened and has a flatter stellar initial mass function (which represents the distribution of stellar masses at the time of starbirth), containing most of the earliest O stars and several near-zero-age main-sequence stars, suggesting a higher proportion of massive stars. The calculated Lyman ionising flux of the earliest stars matches the observed H\,{\sc ii} region emission, and star formation in LH 10 may have been triggered by massive stars in LH 9. UV studies further demonstrate that recent star formation is not confined to star clusters; \citet{2005MNRAS.357..645A} showed that intercluster regions in the LMC can contain young stellar populations comparable to those found within star clusters as seen in a sample of far-UV stars within a $\sim3'$ field near the edge of the LMC bar, between the bar's centre and 30 Doradus. Together, these results established that recent star formation in the Magellanic Clouds is spatially heterogeneous and strongly influenced by stellar feedback, local environmental conditions and the Milky Way--LMC--SMC interactions.

UV observations have also provided important constraints on the recent star-forma\-tion history of the SMC. Using {\sl Swift}/UVOT data, \citet{2017..UVOT..SF..Hagen} identified three prominent peaks of recent star formation in the SMC, occurring at 6--10 Myr, 30--80 Myr and around 400 Myr ago. Their analysis also showed that most of the SMC exhibits the 2175 {\AA} dust bump, which is stronger towards the northeast and weaker in the southwest, and that the extinction curve is generally steeper than that of the Milky Way. More recently, \citet{2025..Ludwig}, using UVOT UV colour--magnitude diagrams (CMDs), reported 522 candidate stripped stars in the LMC and 298 in the SMC. These systems provide valuable laboratories for studying binary evolution and represent an important population linking recent star formation to the progenitors of compact remnants and transient phenomena.

The kinematics of young stellar populations provide important clues to the dynamical evolution of the Magellanic System and the role of galaxy interactions in shaping recent star formation. Using UV observations in combination with astrometric measurements, recent studies have identified signatures of complex stellar motions across the Magellanic Clouds. \citet{2023AA...672A.161M} presented photometry and astrometry for 113 star clusters in the LMC and SMC using {\sl HST} data, providing differential-reddening maps and homogeneous CMDs. We detected extended main-sequence turn-offs in multiple clusters, including previously unreported cases, and showed that their morphology depends on cluster age. Proper-motion measurements allow separation of cluster members from field stars and reveal that young SMC stars exhibit elongated motions towards the LMC, providing new evidence of a kinematic connection between the two galaxies. 

\citet{2024..Hota..SMC} used approximately 62,900 probable SMC far-UV stars to reveal important spatial and kinematic properties of the SMC’s young populations. Far-UV stars are mainly distributed along the main sequence, giant and subgiant branches in CMDs, with about 2200 stars brighter than 15 mag in the far-UV (corresponding to masses $> 8 M_\odot$). The stellar surface-density distribution is notably clumpy, tracing key morphological features such as the broken bar, the inner Wing and a shell-like structure. Population-dependent density maps highlight distinct spatial patterns: Young 2 (main-sequence stars with ages of 50--400 Myr) closely resembles the overall density, whereas Young 1 (main-sequence stars younger than 50 Myr) and blue-loop stars are concentrated at the northeast and southwest of the bar, respectively. The densest regions of Young 3 (main-sequence stars with ages up to 1--2 Gyr) are located in the northeast and in the central part of the bar. More details of these populations can be found in \citet{2021..Gaia} and \citet{2024..Hota..SMC}. The kinematic properties of all these populations suggest that all populations show two kinematically distinct subpopulations. Young 2 and Young 3 show kinematically cool, less dispersed (with a narrower width in the proper-motion distribution) and hot, highly dispersed (with a broad proper-motion distribution) subpopulations, whereas Young 1 and blue-loop stars exhibit two kinematically distinct cool populations. The kinematic analysis of this study points to a specific perturbation for stars younger than $\sim150$ Myr in proper motion along the Right Ascension direction (PMRA), with no significant perturbation in the proper motion along the declination direction (PMDec). This can place useful constraints on the details of the most recent LMC--SMC interaction. 

Using UVIT data, \citet{2024..Hota..SMC..Shell} studied the far-UV sources in the northeastern outskirts of the SMC, the Shell region, which is likely a tidally influenced region. They found a few far-UV sources younger than 7 Myr, with the majority of star formation having occurred between 60 and 300 Myr ago. The spatial distribution of far-UV stars is irregular and clumpy in this Shell region and traces two distinct morphological structures, an arm-like and an arc-like feature. The far-UV stellar distribution exhibits an outwards radial gradient across the SMC Shell, indicating that the inner SMC extends to a radius of roughly $\sim2^{\circ}.2$. Beyond this boundary lies the arm-like structure, while the arc-like feature spans from the inner regions outwards into the SMC’s periphery. The proper motions and velocity dispersion of young far-UV stars, derived from their proper-motion distribution, are consistent with those of the SMC's main body, indicating no significant signs of tidal perturbation or disruption in this region. No evidence of propagating star formation is detected in the SMC's Shell region. Instead, most areas experienced two distinct episodes between 40 and 400 Myr ago: one at $\sim260$ Myr, likely associated with the Magellanic Clouds' most recent interaction, and a younger $\sim60$ Myr-old episode that may correspond to the SMC’s pericentric passage around the Milky Way.

Taken together, these studies demonstrate that the young stellar populations in the Magellanic Clouds retain clear dynamical signatures of their recent interaction history. The spatial distributions, proper motions and age-dependent kinematic properties of UV-bright stars reveal how tidal encounters between the LMC, SMC and Milky Way have shaped recent star formation and the present-day structure of the Magellanic System. UV-selected stellar populations therefore provide a powerful means of linking recent star formation to the broader dynamical evolution of these interacting galaxies.

\subsection{Hierarchical Star Formation}

Stars can appear either as isolated field stars or as members of larger assemblies, such as clusters, stellar associations or extended complexes that span areas as large as the size of a galaxy. Star formation is initiated when dense regions within molecular clouds collapse under their own gravity. The ISM exhibits hierarchical structures across a wide range of spatial scales \citep{1993..Elmegreen}, with giant molecular clouds containing increasingly smaller and denser substructures in a fractal-like arrangement. This organisation is influenced by supersonic turbulence, magnetic fields and gravity \citep{1996..Elmegreen..Efremov}. Young stars, particularly those younger than 100 Myr, are distributed in patterns that reflect the ISM’s hierarchical structure \citep{2015..Gouliermis..HSF..CT}. Observations and simulations both suggest that stellar groupings follow the ISM’s fractal distribution \citep{2001..Elmegreen..&..Elmegreen,2017..Vazquez}, a phenomenon often described as hierarchical, fractal or scale-free star formation \citep{1995..Larson,1998...Bate}.

\begin{figure}[t!]
    \centering
    \includegraphics[width=\textwidth]{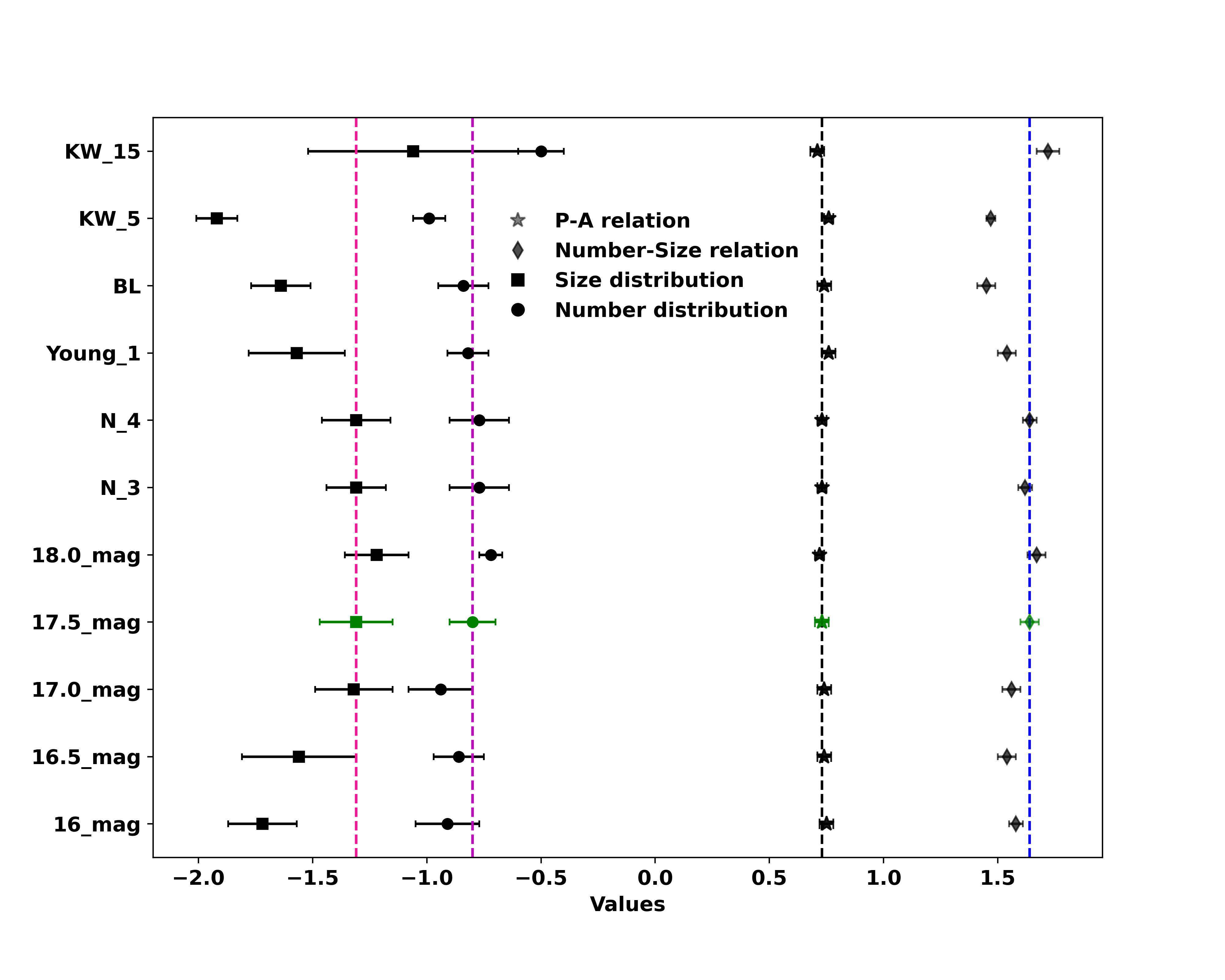}
    \caption{Comparison of the power-law slopes obtained from the number--size, perimeter--area and size-distribution relations under different constraints, including variations in kernel width (KW), $N_\mathrm{min}$, magnitude cutoff and stellar-population composition. The dashed lines indicate the slopes derived in this study for the sample with a far-UV magnitude cutoff of 17.5 and $N_\mathrm{min}=5$ (KW of 10 pc) \citep{2025..Hota..HSF}.}
    \label{fig:Sec6_fig1}
\end{figure}

\citet{2025..Hota..HSF} investigated the star formation traced by young massive stars (5--$40\,M_\odot$) younger than 150 Myr using the catalogue of \citet{2024..Hota..SMC}. They identified 236 young stellar structures composed of bright far-UV stars with high completeness. For each structure, parameters such as size, area, perimeter, surface density and stellar content were measured, with sizes spanning from a few to several hundred parsecs. The perimeter--area relation gives a fractal dimension of $D_p = 1.46 \pm 0.04$, whereas the 2D fractal dimensions from the number--size relation and the size distribution are $D_2 = 1.64 \pm 0.03$ and $D_2 = 1.31 \pm 0.16$, respectively, both consistent within 2$\sigma$. The number distribution between 30 and 1000 follows a slope of $-0.8 \pm 0.1$, close to the expected theoretical value of $-1$ \citep{1996..Elmegreen..HSF,2018...Sun..SMC,2022MNRAS..Miller..LMC}. The surface-density distribution is well described by a lognormal profile, and the results remain robust under different selection criteria. These structures display an hierarchical arrangement similar to that of the ISM, tracing the fractal nature of star formation up to 200 Myr. Figure \ref{fig:Sec6_fig1} presents the slopes of the perimeter--area relation, the number--size relation, the size and number distributions for varying conditions for the far-UV massive stars ($<150$ Myr). This indicates that in the SMC, both the Young 1 and blue-loop populations, which are younger than 200 Myr, exhibit hierarchical patterns. 

A comparison of results obtained using different kernel widths shows that smaller kernel widths yield a larger number of detected structures and, consequently, higher fractal dimensions---although they may also include more noise. By contrast, larger kernel widths produce fewer structures and lower fractal dimensions, implying a loss of genuine substructures. In addition, the fractal dimension decreases with increasing (fainter) magnitude, suggesting that the degree of hierarchy reduces with age. However, varying the minimum number cut-offs does not significantly affect the results. Table~\ref{tab:sum_HSF} presents a summary of the fractal and perimeter--area dimensions reported in different studies. It shows that the far-UV-derived massive young stellar structures exhibit dimensions comparable to those obtained for the near-infrared-identified young, low-mass stars in the SMC. This similarity suggests that the clustering behaviour of young stars is largely independent of stellar mass. It further demonstrates that these dimensions remain consistent across both local regions and the entire SMC, implying that hierarchical patterns persist on both small and large spatial scales. Moreover, despite the differences in galaxy mass of the Milky Way, LMC and SMC, the hierarchical nature of star formation seems a shared characteristic across all three systems.

\begin{table}[t!]
    \centering
    \caption{Summary of estimated fractal and perimeter--area dimensions in different studies.}
    \begin{tabular}{p{0.30\textwidth} p{0.15\textwidth} p{0.15\textwidth} p{0.28\textwidth}}
        \hline
        Study & 2D fractal dimension ($D_2$) & Perimeter--area dimension ($D_P$) & Comments \\
        \hline
        \citet{2025..Hota..HSF} & 1.31--1.64 & 1.46 & UVIT far-UV data;\\ & & & massive young SMC stars \\
        \citet{2022MNRAS..Miller..LMC} & 1.50--1.61 & 1.44 & VMC near-infrared data;\\ & & & entire LMC \\
        \citet{2018...Sun..SMC} & 1.40--1.44 & 1.44 & VMC near-infrared data;\\ & & & entire SMC \\
        \citet{2017..Sun..LMC..Bar} & 1.4--1.6 & -- & VMC near-infrared data;\\ & & & LMC bar \\
        \citet{2017...Sun..30} & 1.3--1.9 & -- & VMC near-infrared data;\\ & & & 30 Doradus (LMC) \\
        \citet{1990..Dickman,2010..Julia} & -- & 1.3--1.6 & Milky Way ISM \\
        \citet{1999..SMC..Stanimirovic..SMC..GLXstr,2000..Stanimirovic} & 1.4--1.5 &  & Gas and dust, SMC \\
        \citet{2014..Gouuliermis..NGC346} & 1.2--1.4 &  & NGC 346 star-forming \mbox{region}, SMC \\
        \citet{2020..SF..regions..MW} & 1.1--1.5 &  & Milky Way star-forming \mbox{regions} \\
        \hline
    \end{tabular}
    \label{tab:sum_HSF}
\end{table}

\subsection{Implications for Stellar Activity and Transient Phenomena}

The Magellanic Clouds host a rich population of UV-bright evolved massive stars that represent key stages in the evolutionary pathways leading to transient phenomena. Their proximity and low metallicity allow individual systems to be studied in detail, providing important constraints on the physical processes governing mass loss, binary interaction and stellar feedback. UV observations are particularly sensitive to hot stellar atmospheres and circumstellar environments, making them a powerful tool for identifying and characterising populations that are often associated with transient progenitors, including Wolf--Rayet stars, stripped stars, high-mass X-ray binaries and luminous blue variables (LBVs).

Recent UV surveys have revealed stellar populations that are directly relevant to transient studies. Using {\sl Swift}/UVOT observations, \citet{2025..Ludwig} identified 522 candidate stripped stars in the LMC and 298 in the SMC. Such stars are thought to originate through binary mass transfer and represent important evolutionary channels leading to stripped-envelope SNe and compact-object binaries \citep{2018A&A...615A..78G, 2020ApJ...904...56G}. Their strong UV excesses make them considerably easier to identify in UV observations than in optical surveys.

UV observations have also played an important role in the study of evolved massive stars that are linked to transient progenitors. Spectroscopic investigations of Wolf--Rayet stars and other hot massive stars in the Magellanic Clouds have provided constraints on mass-loss processes, binary interactions, and the final stages of massive-star evolution \citep{2003IAUS..212...47C,2014AJ....148...64S,2024A&A...689A.157S}. These processes strongly influence the nature of the compact remnants and explosive events produced at the end of stellar evolution. Because radiatively driven winds depend sensitively on metallicity, the Magellanic Clouds provide an important testbed for understanding how mass-loss rates, binary evolution pathways and the production of transient progenitors differ from those observed in the Milky Way.

In addition to progenitor populations, UV observations of SN remnants have provided diagnostics of shock-heated gas and the interaction between stellar ejecta and the surrounding ISM \citep{1982AdSpR...2i.145R}. Such studies complement investigations of young stellar populations by tracing the long-term consequences of stellar feedback and stellar explosions.

These results demonstrate that UV observations contribute not only to our understanding of recent star formation but also to the identification of transient progenitors, the evolution of massive stars and the impact of stellar explosions on their galactic environments, particularly in the low-metallicity environments represented by the Magellanic Clouds.
% Sipra Hota

\section{Outlook}

\subsection{Photometric and spectroscopic perspectives for future UV astronomy}

Throughout this review, UV diagnostics have been discussed primarily in terms of the physical information encoded in emission lines and continua. However, it is important to distinguish between the complementary roles of photometric and spectroscopic observations in extracting that information. Although both approaches probe the same underlying phenomena, they address different scientific questions and impose different requirements on future instrumentation.

UV photometry is particularly powerful for detecting variability and transient phenomena. Wide-field imaging surveys can identify stellar flares, monitor accretion variability in young stellar objects, trace star-forming regions in nearby galaxies and discover rapidly evolving transients such as SN SBO. The principal strengths of photometric observations are their sensitivity, survey speed and temporal coverage, which enable large populations of sources to be monitored simultaneously and the discovery of rare events in real time.

Spectroscopy, by contrast, provides the physical diagnostics needed to interpret these phenomena. Measurements of line strengths, profiles and ratios yield temperatures, densities, abundances, velocity fields and mass-loss rates that cannot generally be recovered from photometry alone. Chromospheric diagnostics such as Mg,{\sc ii}, transition-region tracers including C{\sc iv}, Si{\sc iv} and N{\sc v}, and wind-sensitive resonance lines in massive stars all require spectroscopic observations. Similarly, understanding the energetics of stellar flares, accretion shocks, stellar winds and SN SBO depends critically on spectroscopic measurements that reveal the physical state of the emitting plasma.

Future UV facilities will therefore benefit from maintaining a balance between these complementary capabilities. Wide-field photometric missions, such as {\sl ULTRASAT}, will be particularly effective at identifying transient phenomena and characterising variability across stellar populations. At the same time, the eventual retirement of the {\sl HST} highlights the need for new UV spectroscopic facilities capable of providing high-sensitivity observations over a broad wavelength range. Proposed missions such as {\sl CETUS} \citep{2019arXiv190910437H}, {\sl UVEX} \citep{2021arXiv211115608K}, {\sl CASTOR} \citep{2019clrp.2020...18C,2025JATIS..11d2202C} and related concepts would extend the legacy of the {\sl HST} by enabling detailed investigations of stellar atmospheres, magnetic activity, accretion processes, stellar winds and low-metallicity stellar populations. Among these three proposed facilities, {\sl UVEX} is unusual among survey missions because it includes spectroscopic capability, whereas {\sl CETUS} was conceived from the outset as a spectroscopy-driven mission. {\sl CASTOR} occupies an intermediate position with both imaging and spectroscopic modes.

Ultimately, the greatest scientific return will arise from combining these approaches. Photometric surveys reveal where and when UV variability occurs, whereas spectroscopy determines the physical mechanisms responsible for that variability. Future progress in understanding stellar activity, transient phenomena and their connection to galactic evolution will therefore depend on the continued availability of both capabilities within the UV astronomy landscape.

\subsection{Outstanding questions}
\label{sec8}

UV observations remain among the most incisive yet least exploited tools for probing transient and variable astrophysical phenomena. Throughout this review, we have shown how the UV regime provides unique access to the physical processes that connect the microscopic physics of stellar atmospheres to macroscopic phenomena such as flares, mass loss, SNe and the collective behaviour of stellar populations in galaxies. Yet despite seven decades of UV astronomy, major questions remain unanswered. Whereas the scarcity of UV data with the required cadence, sensitivity and wavelength coverage is a central limitation, theoretical uncertainties and the inherent complexity of multiscale plasma processes also continue to constrain progress.

A first set of challenges concerns stellar flares and magnetic activity. Although the Sun offers a detailed, spatially resolved example, the partition of flare energy across the spectrum remains incompletely characterised for other stars. In particular, the UV energy budget of flares is poorly constrained: how much of the released magnetic energy is radiated in chromospheric and transition-region lines such as Si\,{\sc iv}, C\,{\sc iv} and N\,{\sc v}, and what fraction emerges as continuum emission from heated photospheric layers? Observations of individual events with sufficient spectral and temporal resolution are rare, and existing data sets often lack simultaneous optical and X-ray coverage. For M dwarfs and other magnetically active stars, this uncertainty directly affects models of planetary irradiation, atmospheric escape and photochemistry. A key open question is thus whether the Sun’s flare–energy relationships can be generalised across spectral types and evolutionary stages, or whether low-mass stars follow fundamentally different scaling laws.

A second broad area of uncertainty lies in line formation and radiative transfer in the UV. Even for well-studied resonance transitions---Mg\,{\sc ii} $h$ and $k$, C\,{\sc ii}, Si\,{\sc iv}, C\,{\sc iv} and He\,{\sc ii} 1640 {\AA}---the mapping between observed profiles and the underlying thermodynamic and magnetic structures remains model-dependent. The non-LTE and scattering-dominated nature of these lines makes them exquisitely sensitive to small perturbations in temperature, density or flow velocity but difficult to invert uniquely. How can we disentangle the effects of opacity, PRD and microturbulent broadening to infer reliable physical conditions? Progress will depend on combining high-resolution spectroscopy with advanced numerical radiative transfer and magnetohydrodynamic models. Improved atomic data---particularly for collisional cross-sections and branching ratios---are equally critical.

A third outstanding challenge addresses transient events at high energies and high luminosities, notably SN SBO and the earliest phases of explosive transients. Despite the theoretical expectation that the first photons to escape a stellar envelope appear in the far-UV or soft X-rays, only a handful of direct detections exist. The scarcity of these observations reflects both the short-lived nature of SBOs---minutes to hours---and the limited sky coverage of past UV missions. The coming decade(s), with missions such as {\sl ULTRASAT} and proposed wide-field UV imagers, may finally provide the continuous, high-cadence monitoring needed to capture these fleeting events routinely. Such data would not only constrain progenitor radii and circumstellar structures but also bridge the long-standing observational gap between late-time optical light curves and early theoretical models of explosion physics.

A fourth line of inquiry concerns stellar activity cycles and rotation--activity relations. Although large optical and X-ray surveys have revealed empirical correlations with Rossby number and rotation period, their UV counterparts remain sparse. Does the same two-regime (saturated/unsaturated) behaviour hold in the UV, and how does the UV emission evolve with stellar age and metallicity? Establishing robust relations between UV flux, rotation and convection would enable a unified dynamo framework applicable from the youngest PMS stars to the oldest solar analogues. Multiwavelength surveys that combine photometry, spectroscopy and time-domain coverage are essential to extend these correlations and test their universality.

Beyond individual stars, the extragalactic frontier of UV diagnostics poses its own set of questions. Far-UV spectroscopy of massive stars in the Magellanic Clouds has demonstrated the metallicity dependence of radiatively driven winds, yet it remains unclear whether these relations extrapolate cleanly to the low-metallicity, high-redshift environments of early galaxies. How do massive-star winds, flare activity and feedback processes scale with metallicity, and how do they influence the integrated UV spectra used to interpret cosmic reionisation? Similarly, the hierarchical structures revealed by UV imaging of nearby star-forming regions raise fundamental questions about the universality of star-formation scaling laws and the role of feedback in shaping interstellar environments.

Finally, there is a broad, cross-cutting question concerning instrumentation and access. The field’s progress has historically been constrained by the limited lifetime of UV-capable satellites and the absence of ground-based alternatives. Sustained, high-sensitivity wide-field UV monitoring---coupled with flexible follow-up capabilities---is crucial for the next generation of transient studies. Coordinated observations with optical, infrared and X-ray facilities, alongside advances in theoretical modelling and machine learning-based event classification, will be essential to unlock the full diagnostic potential of the UV regime.

In summary, the UV domain provides a uniquely sensitive diagnostic of dynamic astrophysical processes, tracing the transport of energy from stellar interiors through chromospheres, transition regions and winds, and ultimately into the surrounding ISM. By examining UV diagnostics across the spectrum of transient behaviour---from the microphysics of magnetic reconnection in stellar flares to the macroscopic shocks of SNe and the collective star-formation patterns of galaxies---in this review we have aimed to illustrate how a single, specific wavelength regime can unite the physics of stars, planets and galaxies. The synergy among solar, stellar and extragalactic UV studies is now clearer than ever: the same resonance lines that trace chromospheric heating in the solar atmosphere illuminate the earliest light from exploding stars and shape the spectra of distant galaxies. As we move into an era of rapid, wide-field transient discovery, the UV will remain indispensable---not only as a tracer of energy release and variability, but as a means of connecting the smallest and largest scales of cosmic evolution. Indeed, as we enter a new era of time-domain and multi-messenger astronomy, the UV remains the bridge between the hot, dynamic microphysics of stellar surfaces and the luminous, evolving Universe beyond.

\section*{Acknowledgements}
This article is part of a collection of reviews resulting from the April 2025 Workshop on `Space-based stellar astrophysics in the UV' organised at the International Space Science Institute--Beijing. We are grateful to the institute's staff for excellent organisational and generous financial support. Richard de Grijs acknowledges financial support in the form of a Distinguished Senior Fellowship from the President’s International Fellowship Initiative (PIFI), Chinese Academy of Sciences (Grant No. 2025PVA0007). Ana I. G\'omez de Castro acknowledges support from the Spanish Ministry of Science, Innovation and Universities through Research Grant PID2020-116726RB-I00.

\par\noindent
{\bf Conflicts of interests:} The authors declare no competing conflicts of interest of any kind. No financial support was received beyond that disclosed in the acknowledgements.

\bibliography{uvdiagnostics} 
%\bibliographystyle{sn-basic}
% \printbibliography
\end{document}